\documentclass[preprint,12pt]{elsarticle}

\usepackage{amsmath,amssymb,amsfonts}
\usepackage{graphicx}
\usepackage{booktabs}
\usepackage{multirow}
\usepackage{algorithm}
\usepackage{algorithmic}
\usepackage{xcolor}
\usepackage{subcaption}
\usepackage{bm}
\usepackage{siunitx}
\usepackage{enumitem}
\usepackage{listings}
\usepackage{url}
\usepackage{xurl}
\usepackage[margin=0.75in]{geometry}
\usepackage[section]{placeins}
\usepackage[colorlinks=true,linkcolor=blue,citecolor=blue,urlcolor=blue]{hyperref}
\biboptions{numbers,sort&compress}
\newcommand{\bs}[1]{\boldsymbol{#1}}
\newcommand{\tr}{\operatorname{tr}}

\newcommand{\sym}{\operatorname{sym}}

\newcommand{\macaulay}[1]{\langle #1 \rangle}

\newcommand{\td}{\,\mathrm{d}}
\newcommand{\eps}{\varepsilon}
\newcommand{\sig}{\sigma}
\newcommand{\GC}{G_c}
\newcommand{\lz}{\ell_0}

\begin{document}
\begin{frontmatter}

\title{A matrix-free, differentiable PyTorch solver for phase-field fracture: Formulation, benchmarks, and inverse analysis}

\author[inst1,inst2]{Allamaprabhu Ani}
\author[inst3]{Jean-Fran\c{c}ois Molinari}
\author[inst4]{Ghatu Subhash}
\author[inst1,inst2]{Sathiskumar Anusuya Ponnusami\corref{cor1}}
\ead{s.a.ponnusami@qmul.ac.uk}
\cortext[cor1]{Corresponding author}

\address[inst1]{Department of Engineering, City St. George's, University of London, London, United Kingdom}
\address[inst2]{School of Engineering and Materials Science, Queen Mary University of London, London, United Kingdom}
\address[inst3]{Institute of Civil Engineering, \'{E}cole Polytechnique F\'{e}d\'{e}rale de Lausanne (EPFL), Switzerland}
\address[inst4]{Department of Mechanical and Aerospace Engineering, University of Florida, Gainesville, USA}

\begin{abstract}
A matrix-free, open-source PyTorch solver is presented for phase-field
fracture, designed to run on central processing units (CPUs) and graphics
processing units (GPUs) without custom compiled extensions. In the
explicit dynamic pathway, finite-element operations are formulated as
element-wise tensor contractions with scatter-based accumulation,
removing global sparse mechanics-stiffness assembly from the core
time-stepping loop. Both Ambrosio--Tortorelli regularisations (AT1 and
AT2), multiple energy decompositions (spectral, volumetric-deviatoric,
and star-convex), and plane strain or plane stress assumptions are
supported. The explicit mechanics kernels are compatible with PyTorch's
automatic differentiation engine (autograd), while the implicit,
bound-constrained damage solve is wrapped in a custom backward rule. This
rule implements implicit differentiation through the conjugate-gradient
(CG) linear solve and keeps memory independent of the internal CG
iteration count. The same implementation runs unmodified across macOS,
Linux, and Windows, and has been run on meshes of order $10^6$ nodes on a
single NVIDIA A100 GPU. The solver is compared against four dynamic
fracture cases (straight crack propagation, shear-induced kinking,
dynamic branching, and crack-hole interaction in perforated plates) and
two quasi-static cases (single-edge notched tension and a notched-holed
plate). As a differentiability demonstration, the scalar fracture energy
$\GC$ is recovered from observed crack patterns using PyTorch gradients
through the forward solve and limited-memory Broyden--Fletcher--Goldfarb--Shanno
(L-BFGS) optimisation. Recovery of $\GC$ with relative error below
$10^{-3}$ is achieved after only three accepted L-BFGS states for glass
and two for alumina; further optimiser iterations are retained as a
stability check. The resulting implementation is easier to extend and
combine with differentiable optimisation and machine-learning components.
\end{abstract}


\end{frontmatter}

\clearpage

\section{Introduction}
\label{sec:introduction}

The phase-field approach to fracture replaces the sharp crack topology with a
continuous damage field $d \in [0,1]$ that evolves according to a variational
energy minimisation principle~\cite{francfort1998revisiting,
bourdin2000numerical}. Since no explicit crack tracking, re-meshing, or
enrichment functions are required, the method handles complex crack
phenomena (initiation, branching, merging, and curving) within a standard
finite element framework. This generality has led to its adoption across a wide
range of fracture problems~\cite{ambati2015review, wu2020phase}.

Dynamic phase-field fracture, in which inertial effects play a central role,
has received significant attention since the work of
Borden et al.~\cite{borden2012phase}, who demonstrated crack branching and
the Kalthoff--Winkler impact test using isogeometric analysis (IGA)
with higher-order non-uniform rational B-spline (NURBS) basis
functions. Subsequent contributions include the work
of Hofacker and Miehe~\cite{hofacker2013phase} on the operator-split
dynamic phase-field formulation,
Li et al.~\cite{li2016gradient} on gradient-enhanced damage models, and
Bleyer et al.~\cite{bleyer2017dynamic} on velocity-toughening mechanisms.
Many of these formulations are implemented with implicit or
semi-implicit time integration and assembled sparse matrices. This is a
natural choice for high-order finite-element and IGA discretisations,
where accurate crack-surface regularisation and stable nonlinear solves
are prioritised. Explicit dynamics admits a different computational
route. Local element-wise updates combined with a lumped mass matrix
allow the mechanical update to be written in matrix-free form, avoiding
global sparse assembly in the time-stepping loop
\cite{dewitt2020prisms, davydov2020matrixfreehyperelastic}. In implicit
quasi-static problems, by contrast, the global linear solve remains the
main computational bottleneck even when matrix--vector products are
evaluated matrix-free. Strong preconditioners such as multigrid methods
are typically required in that setting.

Linear-element explicit dynamics with matrix-free operators provide
an efficient path for GPU execution and end-to-end simulation under
PyTorch's automatic differentiation engine (autograd). The
formulation relies on highly parallel tensor operations that map well
to GPU architectures. Because the explicit forward pass is expressed
using standard PyTorch tensor primitives, gradients flow through that
path without handwritten backward kernels. The resulting
implementation is easier to extend or combine with differentiable
optimisation and machine learning components.

Differentiable simulation~\cite{hu2020difftaichi, heiden2021neuralsim}
builds the forward model from operations whose derivatives are known
to the framework, so gradients of any output with respect to any
input flow automatically through the entire algorithm. This enables
gradient-based optimisation, sensitivity analysis, and the training
of neural networks with physics-informed losses, reducing reliance on
finite-difference sensitivities or separately derived adjoint solvers.

Differentiable finite element method (FEM) solvers have been developed
in JAX (JAX-FEM~\cite{xue2023jax}) and
Taichi~\cite{hu2020difftaichi}. Differentiable computational fluid
dynamics (CFD) codes, such as JAX-Fluids~\cite{bezgin2023jaxfluids},
have also been used to differentiate discrete numerical algorithms,
rather than only partial differential equation (PDE) residuals. In the
present work, this idea is applied to explicit dynamic phase-field
fracture in PyTorch.

This setting adds two fracture-specific complications. First, crack
growth is irreversible, so the damage history must be carried through
the computation. Second, the damage update involves a bound-constrained
linear solve. The present implementation handles the forward problem
with matrix-free scatter-based finite-element operators. The damage
gradient is computed with a constant-memory implicit backward rule
around one matrix-free conjugate gradient (CG) solve. The explicit
momentum update is simpler because a lumped mass matrix reduces the
update to diagonal scaling and does not require an adjoint solve. Full
details are given in Section~\ref{sec:implementation}.

The primary advantage of a differentiable forward model is most evident
in inverse problems, where the goal is to recover unknown material
parameters or constitutive fields from observed data. Such problems are
common in experimental fracture mechanics. In dynamic plasticity, for
example, Guo et al.~\cite{guo2014inverse} identify the rate-sensitivity
parameter of a Johnson--Cook model on Ti-6Al-4V from direct-impact
Kolsky bar strain-gauge traces by repeated runs of a commercial finite
element code over a feasible interval. In quasi-static phase-field
fracture, Kosin et al.~\cite{kosin2024parameter} simultaneously
calibrate boundary conditions, Poisson's ratio, fracture energy and
internal length from digital image correlation (DIC) measurements via
integrated DIC. Bayesian frameworks have been developed for scalar
parameters~\cite{wu2021parameter} and for full-field crack
observations~\cite{stanic2026probabilistic}. Gao and
Yoshinaga~\cite{gao2023inverse} recover \emph{spatially
inhomogeneous} fracture toughness fields from observed crack paths
using a phase-field forward model.

These studies often obtain parameter sensitivities by finite
differencing the forward solver or by sampling-based Bayesian
inference. Finite differencing requires one or more additional
simulations per parameter and scales poorly to high-dimensional
problems. Bayesian inference is robust to noise but computationally
expensive. A differentiable forward solver complements both approaches.
A single backward pass computes the full gradient at a cost independent
of the parameter dimension, which makes high-dimensional and
field-valued inverse problems more tractable.

A complementary line of work replaces the forward solver with a
neural-network surrogate. Physics-informed neural networks (PINNs) have
been applied to quasi-static phase-field fracture by
Manav et al.~\cite{manav2024pinnpf}, who report that the sharp damage
gradients at a propagating crack tip remain difficult to represent with
collocation-based residual losses and require specialised training
schedules. The approach taken here keeps the FEM discretisation of the
physics and uses automatic differentiation for gradient extraction. It
therefore avoids replacing the fracture solve with a neural surrogate,
while still allowing the solver to be coupled directly to neural-network
training loops. Because the implementation is built entirely on PyTorch
tensor operations, the solver and any downstream neural network can
share a single autograd graph. This avoids file-based exchange with
traditional C++ solvers and avoids a framework boundary between JAX and
PyTorch.

The solver operates as follows:
\begin{enumerate}[leftmargin=*,nosep]
  \item All finite element operations (strain, stress, internal force) are
        computed element-wise using vectorised tensor operations.
  \item Global assembly is performed by an indexed scatter
        accumulation (atomic accumulation by node index), a
        differentiable primitive in PyTorch.
  \item The momentum equation is advanced explicitly with velocity-Verlet
        and lumped mass (diagonal inversion), requiring no linear solve.
  \item The phase-field equation is solved by a preconditioned conjugate
        gradient method, which itself is composed of differentiable
        operations (matrix-vector products via scatter, inner products,
        scalar updates).
\end{enumerate}
The explicit tensor kernels are differentiated by PyTorch autograd, and
the CG damage solve is differentiated through the implicit rule
described later. The demonstrated inverse application is detailed in
Section~\ref{sec:differentiability}.

The contributions of this paper are:
\begin{enumerate}[leftmargin=*,nosep]
  \item A matrix-free implementation of explicit dynamic phase-field
        fracture in PyTorch, using standard tensor operations in the
        time-stepping loop (no author-written CUDA kernels and no
        sparse matrices in the explicit operator application).
        Both Ambrosio--Tortorelli phase-field models (AT1 and AT2)
        are supported, with spectral, volumetric-deviatoric and
        star-convex~\cite{kumar2020revisiting} energy decompositions,
        under plane strain or plane stress.
  \item Benchmark evidence across two benchmark families.
        First, two AT2/spectral/plane-strain dynamic checks using
        Borden et al.'s glass and steel benchmark parameters:
        a straight mode-I propagation variant and the
        Kalthoff--Winkler impact setup~\cite{borden2012phase}, together
        with an AT1/Amor dynamic crack-branching check based on the
        COMSOL application example~\cite{comsol_dynamic_branching64}.
        Second, a perforated polymethyl methacrylate (PMMA)
        benchmark is included in AT1/Amor/plane-stress form, using
        representative hole configurations to probe constrained
        mid-plane propagation, single distant-hole interaction, and
        multiple distant-hole interaction following the crack-hole
        mechanisms studied by Bleyer et al.~\cite{bleyer2017dynamic}.
        In addition, two
        quasi-static checks are reported for single-edge notched
        tension (SENT) and a notched-holed plate
        benchmark~\cite{comsol_holed_plate64}.
  \item A secondary consequence of the pure-tensor design is that the
        explicit tensor kernels are compatible with PyTorch's automatic
        differentiation engine (autograd), while the CG damage solve is
        differentiated through a custom implicit rule.
        This is demonstrated on an L-BFGS scalar $\GC$
        inversion (Section~\ref{sec:differentiability}). The same
        differentiable forward map motivates future higher-dimensional
        inverse and design studies.
  \item An open-source implementation that runs unmodified on
        macOS, Linux, and Windows, on both CPU and NVIDIA GPU
        backends, with no compiled extensions or platform-specific
        build steps.
\end{enumerate}

The remainder of this paper is organised as follows.
Section~\ref{sec:governing} presents the governing equations and the
matrix-free implementation, including the explicit time integration
scheme.
Section~\ref{sec:benchmarks} presents the benchmark validation.
Section~\ref{sec:differentiability} discusses differentiability and its
applications.
Section~\ref{sec:performance} provides performance measurements.
Conclusions are drawn in Section~\ref{sec:conclusions}.

\section{Model and matrix-free implementation}
\label{sec:governing}

\subsection{Continuous formulation}
\label{sec:continuous}

\begin{figure}[htbp]
  \centering
  \includegraphics[width=0.97\linewidth]{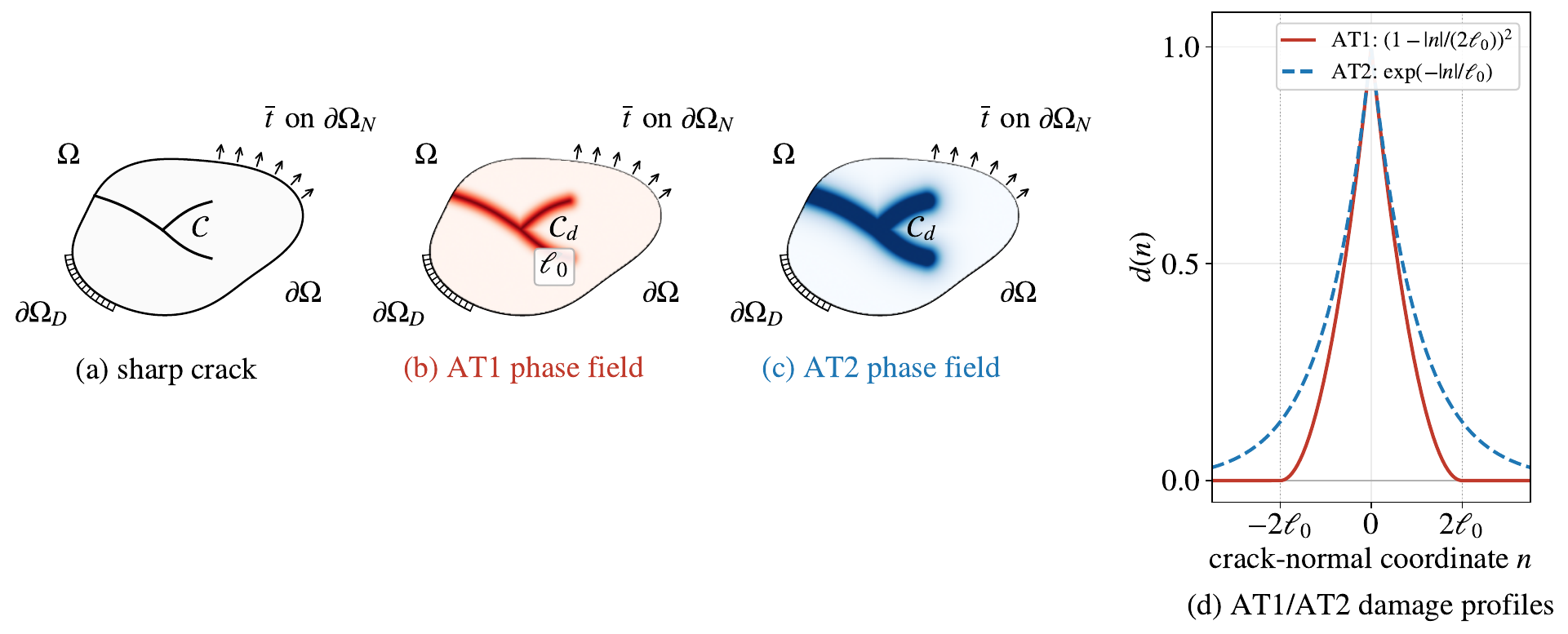}
  \caption{Regularisation of a sharp crack in an arbitrary body
  $\Omega$ with boundary $\partial\Omega = \partial\Omega_D \cup
  \partial\Omega_N$ (Dirichlet support + Neumann traction $\bar{t}$).
  (a)~Classical sharp-interface description: the crack
  $\mathcal{C}$ is a one-dimensional discontinuity emerging from the
  Dirichlet edge. The phase-field representations (b)--(c) replace
  this discontinuity with a continuous damage field
  $d(\bs{x}) \in [0,1]$ supported in a regularised band
  $\mathcal{C}_d = \{\bs{x} : d(\bs{x}) > 0\}$ around the crack
  centreline; the regularisation length $\lz$ is annotated on (b).
  (b)~AT1 band: compact support, the field is identically zero outside
  $|n| \le 2\lz$ along the crack normal $n$. (c)~AT2 band: same
  centreline but an exponential tail extends throughout the body.
  (d)~The 1-D analytical profiles $d(n)$ for the two regularisations:
  AT1 $d(n) = \langle 1 - |n|/(2\lz)\rangle_{+}^{2}$ has compact
  support and a finite elastic threshold; AT2
  $d(n) = \exp(-|n|/\lz)$ has an exponential tail.}
  \label{fig:regularization_schematic}
\end{figure}

A body occupying the domain $\Omega \subset \mathbb{R}^2$ is considered,
described by a displacement field $\bs{u}(\bs{x},t)$ and a phase-field
variable $d(\bs{x},t) \in [0,1]$, with $d=0$ denoting intact material
and $d=1$ a fully developed crack. The discontinuous crack set
$\mathcal{C}$ that would appear in a sharp-interface description of
brittle fracture is replaced here by a continuous regularised band
$\mathcal{C}_d = \{\bs{x} : d(\bs{x}) > 0\}$ of characteristic width
$2\lz$, where $\lz$ is a regularisation length introduced below
(Fig.~\ref{fig:regularization_schematic}). Following the variational formulation of brittle
fracture~\cite{francfort1998revisiting, bourdin2000numerical} and its
dynamic extension~\cite{borden2012phase, hofacker2013phase}, the
energy functional contains elastic, fracture, kinetic, and external-work
contributions,
\begin{equation}
  \mathcal{E}(\bs{u}, d) =
    \int_\Omega \psi(\bs{\eps}, d) \td\Omega
    + \int_\Omega \GC \, \gamma(d, \nabla d) \td\Omega
    + \int_\Omega \frac{\rho}{2} |\dot{\bs{u}}|^2 \td\Omega
    - \mathcal{W}_{\text{ext}},
  \label{eq:total_energy}
\end{equation}
where $\bs{\eps} = \sym(\nabla \bs{u})$ is the linearised strain, $\rho$
is the mass density, and $\mathcal{W}_{\text{ext}}$ is the external
work. The crack surface density $\gamma(d, \nabla d)$ depends on the
chosen phase-field model. The Ambrosio--Tortorelli 2 (AT2)
model~\cite{bourdin2000numerical} uses a quadratic local term,
$\gamma_{\text{AT2}} = d^2 / (2\lz) + (\lz/2)|\nabla d|^2$, while the
Ambrosio--Tortorelli 1 (AT1) model~\cite{pham2011gradient} uses a linear one,
$\gamma_{\text{AT1}} = (3/8)\big(d/\lz + \lz |\nabla d|^2\big)$, where
$\lz$ is the regularisation length. AT1 possesses a finite elastic
threshold. Damage nucleates only when the driving force exceeds
$\mathcal{H}_{\text{crit}} = 3\GC/(16\lz)$, with critical stress
$\sigma_c = \sqrt{3\GC E/(8\lz)}$. The AT2 model has no threshold and damage
initiates immediately under any positive strain energy. Both models
are implemented in the present solver.

To prevent damage growth under compressive loading, the elastic energy
is split into a tensile (degradable) and a compressive (preserved) part.
The solver implements both the spectral split of
Miehe et al.~\cite{miehe2010thermodynamically},
\begin{equation}
  \psi^{\pm}_{\text{spec}}(\bs{\eps})
    = \tfrac{\lambda}{2}\macaulay{\tr(\bs{\eps})}_{\pm}^2
    + \mu\bigl(\macaulay{\eps_1}_{\pm}^2 + \macaulay{\eps_2}_{\pm}^2\bigr),
  \label{eq:spectral_split}
\end{equation}
where $\eps_1, \eps_2$ are the principal strains, and the
volumetric-deviatoric split of Amor et al.~\cite{amor2009regularized},
\begin{equation}
  \psi^{+}_{\text{Amor}}(\bs{\eps})
    = \tfrac{K}{2}\macaulay{\tr(\bs{\eps})}_{+}^2
    + \mu\,\bs{\eps}_{\text{dev}}\!:\!\bs{\eps}_{\text{dev}}, \qquad
  \psi^{-}_{\text{Amor}}(\bs{\eps})
    = \tfrac{K}{2}\macaulay{\tr(\bs{\eps})}_{-}^2,
  \label{eq:amor_split}
\end{equation}
with $K$ the three-dimensional bulk modulus and $\bs{\eps}_{\text{dev}}$
the deviatoric strain. In both decompositions the total degraded elastic
energy is
$\psi(\bs{\eps}, d) = g(d)\,\psi^+(\bs{\eps}) + \psi^-(\bs{\eps})$
with $g(d) = (1-d)^2 + \eta$, $\eta = 10^{-7}$.
For reference-matching quasi-static cases, the implementation can also
use the fully isotropic choice $\psi^+ = \psi$ and $\psi^- = 0$, so that
all elastic strain energy is degraded by the phase field.

\paragraph{Plane assumptions.}
Two-dimensional simulations may be performed under plane strain
($\eps_{zz} = 0$) or plane stress ($\sigma_{zz} = 0$).  The two
assumptions differ only in the effective first Lam\'{e} parameter. The
subscripts PE and PS denote plane strain and plane stress, respectively:
$\lambda_{\text{PE}} = E\nu/[(1{+}\nu)(1{-}2\nu)]$ versus
$\lambda_{\text{PS}} = E\nu/(1{-}\nu^2)$. The distinction propagates
through the constitutive operators. The volumetric-deviatoric
split~\eqref{eq:amor_split} and the star-convex variant of
Kumar et al.~\cite{kumar2020revisiting} both retain the
three-dimensional bulk modulus $K = E/[3(1{-}2\nu)]$. Under plane
stress, these splits recover the out-of-plane strain as
$\eps_{zz} = -\nu/(1{-}\nu)\,\tr_{2D}(\bs{\eps})$ ensuring that the
volumetric term $K\,\tr(\bs{\eps})$ remains thermodynamically
consistent. In the validation benchmarks reported below, the
Miehe strain-spectral split is used in plane strain, while the
plane-stress PMMA benchmark uses the Amor volumetric-deviatoric split
with this out-of-plane contribution included in the three-dimensional
trace.

\subsection{Strong form and staggered scheme}
\label{sec:staggered_scheme}

The momentum equation is
\begin{equation}
  \rho \ddot{\bs{u}} = \nabla \cdot \bs{\sig} + \bs{b}
  \quad \text{in } \Omega,
  \label{eq:momentum}
\end{equation}
where $\bs{\sig} = g(d)\,\partial\psi^+/\partial\bs{\eps}
+ \partial\psi^-/\partial\bs{\eps}$ is the Cauchy stress. The mechanical
boundary conditions are
\begin{equation}
  \bs{u} = \bar{\bs{u}} \quad \text{on } \partial\Omega_D,
  \qquad
  \bs{\sig}\bs{n} = \bar{\bs{t}} \quad \text{on } \partial\Omega_N .
  \label{eq:mechanical_boundary_conditions}
\end{equation}
Damage
irreversibility is enforced through the history variable
$\mathcal{H}(\bs{x},t) = \max_{\tau \leq t} \psi^+(\bs{\eps}(\bs{x},\tau))$.
For the AT2 model the phase-field equation is the linear
elliptic problem
\begin{equation}
  \left( \frac{\GC}{\lz} + 2\mathcal{H} \right) d
  - \GC \lz \, \Delta d = 2\mathcal{H}
  \quad \text{in } \Omega,
  \label{eq:damage_pde_at2}
\end{equation}
with the natural phase-field boundary condition
\begin{equation}
  \nabla d \cdot \bs{n} = 0
  \quad \text{on } \partial\Omega .
  \label{eq:damage_boundary_condition}
\end{equation}
For AT1 the equation becomes a variational inequality~\cite{pham2011gradient},
\begin{equation}
  2\mathcal{H}\,d - \tfrac{3}{4}\GC\lz\,\Delta d
  \geq 2\mathcal{H} - \tfrac{3\GC}{8\lz},
  \qquad d \geq 0,
  \label{eq:damage_pde_at1}
\end{equation}
with equality on the active set where damage grows and the usual
complementarity condition between the residual and the bound. The
bound constraint $d \geq 0$ is essential and is enforced via a
projected preconditioned conjugate-gradient (CG) solver
(Section~\ref{sec:implementation}).

The coupled mechanical and phase-field problems are solved in a
staggered fashion: at each time step the displacement $\bs{u}_{n+1}$
is advanced explicitly with $d_n$ fixed; the strain energy $\psi^+$
and the history variable $\mathcal{H}_{n+1} = \max(\mathcal{H}_n,
\psi^+_{n+1})$ are then updated; finally
\eqref{eq:damage_pde_at2} (or \eqref{eq:damage_pde_at1} for AT1) is
solved for $d_{n+1}$. For explicit dynamics with a small time step, a
single staggered pass per step is sufficient~\cite{borden2012phase}.

\subsection{Matrix-free gather--compute--scatter discretisation}
\label{sec:implementation}

The solver is implemented in Python using
PyTorch~\cite{paszke2019pytorch}. The central design principle is that
the explicit operator application is matrix-free. Finite-element
operations are expressed as compositions of three primitives,
\emph{gather} (extract element-local data from global arrays),
\emph{element-level computation} (strain, stress, energy), and
\emph{scatter} (accumulate element contributions back into global
arrays). These three primitives are standard differentiable operations
in PyTorch, which is what makes the entire forward pass amenable to
automatic differentiation (Section~\ref{sec:differentiability}).

The domain is discretised with linear triangular (P1) elements
generated by Gmsh~\cite{geuzaine2009gmsh} with graded refinement near
the expected crack path. The nodal coordinates
$\bs{X} \in \mathbb{R}^{N \times 2}$ and the element connectivity
$\bs{T} \in \mathbb{N}^{N_e \times 3}$ are stored as PyTorch tensors.
Because the shape function gradients of P1 elements are constant per
element, the gradient tensor
$\nabla\phi \in \mathbb{R}^{N_e \times 3 \times 2}$ stores one
$(\partial_x \phi_i,\partial_y \phi_i)$ row for each local node,
\begin{equation}
  \nabla\phi^e = \frac{1}{2 A_e}
  \begin{bmatrix}
    y_2 - y_3 & x_3 - x_2 \\
    y_3 - y_1 & x_1 - x_3 \\
    y_1 - y_2 & x_2 - x_1
  \end{bmatrix},
  \label{eq:grad_phi}
\end{equation}
and the area vector ${A} \in \mathbb{R}^{N_e}$ are precomputed once
during initialisation. Every global finite-element (FE) operation then follows the same
gather--compute--scatter pattern. Element-local displacement arrays
$\bs{u}^e$ are obtained by index gathering, element strains and
stresses are computed by vectorised contractions over all $N_e$
elements simultaneously, and the resulting element forces are
scattered back to the global force vector via PyTorch's atomic
indexed accumulation, which provides the required parallel
accumulation on GPUs.

The key observation is that a Krylov solver does not need the entries
of the stiffness matrix. It only needs the product of the stiffness
matrix with a trial vector. In the familiar assembled linear-elastic
form, the internal force is
\begin{equation}
  \bs{f}_{\mathrm{int}}(\bs{u})=\bs{K}\bs{u}.
  \label{eq:linear_force_stiffness}
\end{equation}
In this setting, applying the stiffness matrix to a trial displacement
direction $\bs{p}$ is equivalent to asking the internal-force routine
what force would be generated by the displacement $\bs{p}$. A
matrix-free implementation therefore evaluates the product as
\begin{equation}
  \bs{K}\bs{p}
  =
  \bs{f}_{\mathrm{int}}(\bs{p})
  -
  \bs{f}_{\mathrm{int}}(\bs{0}).
  \label{eq:matrix_free_force_difference}
\end{equation}
Here $\bs{f}_{\mathrm{int}}(\bs{0})$ means the internal force returned
by the same element routine when all nodal displacements are set to
zero, with the same mesh, material state, boundary bookkeeping, and
fixed damage field. In an ideal linear elastic problem with no offset,
$\bs{f}_{\mathrm{int}}(\bs{0})=\bs{0}$, and
Eq.~\eqref{eq:matrix_free_force_difference} reduces to the standard
relation $\bs{K}\bs{p}=\bs{f}_{\mathrm{int}}(\bs{p})$. The subtraction
is kept in the implementation because it removes any constant offset
introduced by constraints, fixed state variables, or numerical
bookkeeping. It also guarantees that the matrix--vector product maps
the zero trial direction to zero, as a linear operator must.

For a nonlinear or state-dependent problem, the same idea is applied to
the tangent operator at the current reference state $\bar{\bs{u}}$,
\begin{equation}
  \bs{K}(\bar{\bs{u}})\bs{p}
  =
  \left.
  \frac{d}{d\epsilon}
  \bs{f}_{\mathrm{int}}(\bar{\bs{u}}+\epsilon\bs{p})
  \right|_{\epsilon=0}.
  \label{eq:matrix_free_tangent_action}
\end{equation}
When the internal force is affine over the current operator
application, this tangent action can be evaluated by the shifted
difference
\begin{equation}
  \bs{K}(\bar{\bs{u}})\bs{p}
  =
  \bs{f}_{\mathrm{int}}(\bar{\bs{u}}+\bs{p})
  -
  \bs{f}_{\mathrm{int}}(\bar{\bs{u}}).
  \label{eq:matrix_free_shifted_difference}
\end{equation}
This is the sense in which the stiffness matrix is represented by its
action on vectors rather than by assembled sparse entries.

The difference from a classical sparse finite-element implementation is
when the weak-form operator is materialised. In an assembled code,
each element contributes a small local stiffness matrix, and these local
matrices are inserted into a global sparse matrix. If $\bs{P}_e$ is the
Boolean extraction matrix that maps global nodal values to element
values, $\bs{u}^e=\bs{P}_e\bs{u}$, the classical assembled operator is
\begin{equation}
  \bs{K}
  =
  \sum_{e=1}^{N_e}
  \bs{P}_e^{T}\bs{K}^{e}\bs{P}_e,
  \qquad
  \bs{K}\bs{p}
  =
  \sum_{e=1}^{N_e}
  \bs{P}_e^{T}\bs{K}^{e}\bs{p}^{e}.
  \label{eq:assembled_fe_operator}
\end{equation}
The first expression explicitly builds the sparse matrix $\bs{K}$.
The second expression is then evaluated by sparse matrix--vector
multiplication or by a sparse direct or iterative solver. The resulting
matrix can be reused, factorised, preconditioned, or passed to mature
sparse linear-algebra libraries. This is often the most convenient
choice for implicit Newton solves and for problems where direct sparse
factorisation is affordable. The cost is that assembly creates a large
global data structure with irregular memory access, and the sparse
matrix must be rebuilt when the tangent changes.

In the matrix-free form, the global matrix is never stored. Each
operator application repeats the element gather, local weak-form
evaluation, and scatter accumulation. The same action is evaluated as
\begin{equation}
  \bs{q}^{e}=\bs{K}^{e}\bs{p}^{e},
  \qquad
  \bs{q}
  =
  \operatorname{scatter}
  \left(
    \left\{\bs{P}_e^{T}\bs{q}^{e}\right\}_{e=1}^{N_e}
  \right),
  \qquad
  \bs{q}=\bs{K}\bs{p}.
  \label{eq:matrix_free_fe_operator}
\end{equation}
Here $\bs{K}^{e}$ need not be stored as a local matrix. In the present
implementation, $\bs{q}^{e}$ is obtained directly from the element
weak-form contraction or from the internal-force difference in
Eq.~\eqref{eq:matrix_free_force_difference}. This is less convenient
for direct solvers because no explicit matrix is available to factorise,
but it is well matched to explicit dynamics and Krylov methods. The
memory footprint is lower, the computation is expressed as batched
tensor operations, and the same code path maps naturally to GPUs and
PyTorch autograd.

\paragraph{Velocity-Verlet integration.}
\label{sec:verlet}

The momentum equation~\eqref{eq:momentum} is integrated using the 
velocity-Verlet scheme. This scheme advances the state from $t_n$ to 
$t_{n+1} = t_n + \Delta t$ in three explicit substeps. These are a displacement
update (the \emph{predictor}), an internal-force evaluation at the new 
displacement that yields the new acceleration (the \emph{force evaluation}), 
and a velocity update (the \emph{corrector}).

\begin{align}
  \bs{u}_{n+1} &= \bs{u}_n + \Delta t \, \bs{v}_n
    + \tfrac{1}{2} \Delta t^2 \, \bs{a}_n, \label{eq:vv_u} \\
  \bs{a}_{n+1} &= \bs{m}^{-1}
    \left( \bs{f}_{\text{ext}} - \bs{f}_{\text{int}}(\bs{u}_{n+1}, d_n) \right), \label{eq:vv_a} \\
  \bs{v}_{n+1} &= \bs{v}_n + \tfrac{1}{2} \Delta t
    \left( \bs{a}_n + \bs{a}_{n+1} \right), \label{eq:vv_v}
\end{align}

where $\bs{m}$ is the lumped (diagonal) mass vector with entries 
$m_i = \rho \sum_{e \ni i} A_e/3$. Its inversion is a trivial 
element-wise division. The stable time step is chosen according to 
the Courant--Friedrichs--Lewy (CFL) condition 
$\Delta t \leq \alpha\,h_{\min}/c_p$, where 
$c_p = \sqrt{(\lambda + 2\mu)/\rho}$ is the dilatational wave speed, 
$h_{\min}$ is the minimum element incircle diameter, and $\alpha = 0.8$ 
is a safety factor.

\subsection{Damage sub-problem}
\label{sec:damage_solver}

The phase-field equation is a symmetric positive-definite linear system 
that is solved at each (subcycled) time step using a preconditioned 
conjugate gradient method. The matrix--vector products follow the same 
gather--compute--scatter pattern as the mechanical solver. For a trial
damage vector $p$, the kernel gathers $p^e$ on each element, computes
the diffusion contribution from $\nabla p_e$ and the reaction
contribution from the consistent mass matrix, and scatters the
element result into the global product $\bs{q}=A p$ without forming
the matrix $A$.

An algebraic multigrid (AMG) preconditioner is employed to accelerate
the damage solve, typically reducing the CG iteration count by a factor
of $5$ to $20$ on fine meshes. The AMG hierarchy is constructed once on
the host by PyAMG~\cite{bell2022pyamg} from a sparse system matrix that
is assembled only temporarily. This matrix exists solely to initialise
the preconditioner and is discarded immediately afterwards. The
subsequent CG iterations do not access it.

The setup cost is therefore a one-off expense amortised over many time
steps. The hierarchy is reused until the damage growth exceeds a
prescribed threshold, at which point it is rebuilt. The AMG operations
are executed on the compute device. On CPU and generic GPUs, the
restriction, prolongation, smoothing, and coarse solve are performed
with PyTorch tensor routines. On NVIDIA hardware, an optional AmgX
library~\cite{naumov2015amgx} backend executes the full preconditioner
using cuSPARSE and cuSOLVER. When rapid damage growth invalidates the
AMG hierarchy, a regime common during active AT1 damage, the solver
falls back to a Jacobi preconditioner with exponential backoff before
retrying AMG.

Bound enforcement ensures that the damage field satisfies the physical
constraints $d \ge 0$ and $d \ge d_{\text{prev}}$ (irreversibility).
The strategy depends on the phase-field model. For the AT2 model, the 
reaction term $(\GC/\lz + 2\mathcal{H})$ is always positive, so the 
unconstrained solution remains non-negative. Irreversibility can thus 
be enforced by simple clamping after the CG solve. 

For the AT1 model, the lower bound $d \ge 0$ must be imposed 
\emph{during} the iterative solve. An unconstrained solve produces 
large negative values of $d$ in regions far from the crack, where 
the reaction term $2\mathcal{H}$ vanishes. These negative values 
propagate to the crack tip through the Laplacian operator, often 
resulting in a fully healed solution ($d \approx 0$) after post-clamping. 
To address this, a projected preconditioned conjugate gradient
(PPCG) method is used. This approach maintains an active set of bound-constrained
nodes. At each iteration, nodes at the lower bound with negative residual 
are frozen, while CG operates only on the free degrees of freedom. 
When a bound becomes inactive, the solver restarts on the updated free set.

This method achieves the same effect as bound-constrained quadratic 
optimisation~\cite{bleyer2017dynamic}, but is implemented entirely 
in PyTorch. It therefore remains compatible with GPU execution and 
end-to-end automatic differentiation. The solver automatically detects 
the model type and switches between the post-clamping (AT2) and 
projected (AT1) strategies without user intervention.

\label{sec:projected_cg}

The damage equation uses a consistent (non-lumped) mass matrix for the
reaction term. A lumped mass produces damage fields that are too
diffuse because the lumping smears the reaction term across the
node patches (see also Bourdin et al.~\cite{bourdin2000numerical}
for the foundational variational damage formulation in which the
$d$-mass term first appears). The element mass entries
$M^e_{ij} = A_e (1 + \delta_{ij})/12$ are evaluated by the same
batched contraction used by the matrix-free damage matvec.

Motivated by the observed benchmark crack speeds, typically below
about $0.6\,c_R$ in the dynamic propagation benchmarks, the solver
optionally subcycles the damage sub-problem. Rather than solving the
phase-field PDE at every explicit step, the solver evaluates it only
every $N_{\text{sub}}$-th step while the mechanics sub-problem advances
normally. The justification is that the damage front propagates at
$\sim 0.6\,c_R$, whereas the explicit time step is limited by the
faster wave speed ($c_p \approx 3 \times (0.6\,c_R)$). The damage field
therefore evolves more slowly than the displacement field. The
continuous speed ratio $c_p/(0.6\,c_R)$ gives a conservative estimate
of the admissible subcycling factor, and the integer value used by the
solver is $N_{\max} = \lfloor c_p/(0.6\,c_R) \rfloor$. A
pre-simulation check computes the elastic wave speeds, the CFL time
step, and the per-material $N_{\max}$ before time integration begins.
PMMA ($\nu = 0.35$) gives a continuous ratio of $3.4$, so the integer
limit is $N_{\max}=3$. All benchmark materials reported here satisfy
$N_{\max} \geq 3$.
The history variable $\mathcal{H}$ is still updated every step, so
damage irreversibility is enforced on the fine mechanical timescale.
\label{sec:subcycling}

The phase-field formulation involves quantities spanning several
orders of magnitude. For example, with the Borden glass parameters,
$\GC \cdot \lz = 7.5 \times 10^{-4}$~N while $E = 32{,}000$~MPa, and
the ratio $\GC/\lz = 0.012$~MPa is six orders of magnitude smaller
than the elastic modulus. In single precision
($\eps_{\text{mach}} \approx 6 \times 10^{-8}$) the damage solver's
residual can stagnate above the convergence tolerance due to loss of
significance in the reaction term. The solver therefore enforces
64-bit arithmetic for all solver-critical operations, and on NVIDIA
hardware PyTorch's automatic mixed-precision (AMP) autocast is
explicitly disabled in the integration loop so that the
double-precision policy is respected end-to-end.

\section{Benchmark validation}
\label{sec:benchmarks}

The solver is benchmarked against two benchmark families. The first is a
set of four dynamic fracture benchmarks ordered by increasing
crack-topology complexity (straight propagation, kinking, branching,
and crack-hole interaction), summarised in
Table~\ref{tab:benchmark_overview}.
Sections~\ref{sec:dsent} and \ref{sec:kalthoff} use Borden et
al.'s material parameters and AT2/spectral/plane-strain formulation; the
Kalthoff--Winkler geometry in Section~\ref{sec:kalthoff} follows
Borden et al.~\cite{borden2012phase} directly.
Section~\ref{sec:borden_branching} uses a dynamic crack-branching
plate in the AT1/Amor/plane-strain setting~\cite{comsol_dynamic_branching64}.
Section~\ref{sec:perforated} follows Bleyer et
al.~\cite{bleyer2017dynamic} on the AT1/Amor/plane-stress setup. The
second is a quasi-static set consisting of
single-edge-notched tension (SENT) and a notched-holed plate
benchmark~\cite{comsol_holed_plate64}. Together, these benchmarks
exercise both phase-field models, both loading regimes, both energy
decompositions, and both plane assumptions implemented in the solver.

Several loading protocols are used across the benchmarks. The B1
dynamic SENT case is displacement-driven and applies a smooth
Dirichlet ramp to the top and bottom boundaries. The total opening is
$0.002$~mm and the ramp duration is $20\,\mu$s. The B3 dynamic
crack-branching case is traction-driven and applies opposing tractions
through a smooth load factor. Here $t$ denotes the elapsed simulation
time and $t_r$ denotes the ramp duration. The load factor is
\begin{equation}
  S(t) =
  \begin{cases}
    0, & t \leq 0,\\
    \tfrac{1}{2}\left[1-\cos\left(\pi t/t_r\right)\right],
      & 0 < t < t_r,\\
    1, & t \geq t_r .
  \end{cases}
  \label{eq:smooth_traction_ramp}
\end{equation}
This smooth onset removes an artificial load jump while preserving the
intended dynamic loading. The retained crack-branching run uses
$t_r=0.05\,\mu$s.

The Kalthoff--Winkler benchmark uses a prescribed impact displacement
that corresponds to a velocity ramp. During the ramp, the impact
velocity increases linearly to the target value $v_0$. It is then held
constant for the rest of the run,
\begin{equation}
  v(t) =
  \begin{cases}
    v_0\,t/t_r, & 0 \leq t < t_r,\\
    v_0, & t \geq t_r .
  \end{cases}
  \label{eq:kalthoff_velocity_ramp}
\end{equation}
with $t_r=1\,\mu$s. The benchmarks following
Bleyer et al.~\cite{bleyer2017dynamic} use a different pre-strain
protocol. In those cases the plate is first brought to a pre-strained
quasi-static equilibrium and is then released at $t=0$. No transient
loading ramp is applied during the dynamic run.

\begin{table}[htb]
\centering
\caption{Benchmark setup: material parameters, discretisation, and loading.
Young's modulus $E$, Poisson's ratio $\nu$,
density $\rho$, Griffith fracture energy $\GC$ in N/mm, and
regularisation length $\lz$ are listed. All simulations use linear
triangular elements with graded refinement and 64-bit arithmetic. The
validation results are summarised in Table~\ref{tab:benchmark_summary}
and detailed in the per-benchmark subsections below.}
\label{tab:benchmark_overview}
\resizebox{\linewidth}{!}{%
\footnotesize
\setlength{\tabcolsep}{4pt}
\begin{tabular}{@{}lllccrrrrrl@{}}
\toprule
ID & Benchmark / Material & Model / Split / Plane &
   $E$ (GPa) & $\nu$ & $\rho$ (kg/m$^3$) & $\GC$ (N/mm) & $\lz$ (mm) &
   Domain (mm) & Nodes & Loading \\
\midrule
B1 & Dynamic SENT, glass &
   AT2 / Spec.\ / Strain &
   $32.0$ & $0.20$ & $2450$ & $0.003$ & $0.50$ &
   $40\!\times\!40$ & 1{,}091 &
   smooth Dirichlet opening ramp \\
B2 & Kalthoff--Winkler, steel &
   AT2 / Spec.\ / Strain &
   $190.0$ & $0.30$ & $8000$ & $22.13$ & $0.195$ &
   $100\!\times\!100$ & 3k to 213k &
   $v = 16.5$~m/s ramp \\
\midrule
B3 & Dynamic crack branching~\cite{comsol_dynamic_branching64} &
   AT1 / Amor / Strain &
   $32.0$ & $0.20$ & $2450$ & $0.003$ & $0.50$ &
   $100\!\times\!40$ & 169k &
   smooth traction ramp \\
B4 & Perforated plate, PMMA~\cite{bleyer2017dynamic} &
   AT1 / Amor / Stress &
   $3.09$ & $0.35$ & $1180$ & $0.300$ & $0.10$ &
   $32\!\times\!16$ & 50k to 214k &
   pre-strain $\Delta U \in \{0.04,0.05\}$~mm + holes \\
\midrule
Q1 & SENT, steel &
   AT2 / Iso.\ / Strain &
   $210.0$ & $0.30$ & $7800$ & $2.70$ & $0.015$ &
   $1\!\times\!1$ & 2{,}423 &
   monotone top-edge displacement \\
Q2 & Notched-holed plate, mortar~\cite{comsol_holed_plate64} &
   AT2 / Iso.\ / Stress &
   $6.0$ & $0.22$ & $2400$ & $2.28$ & $0.25$ &
   $65\!\times\!120$ & 19{,}129 &
   rigid-pin displacement \\
\bottomrule
\end{tabular}}
\end{table}

Meshes are generated with Gmsh using graded refinement, with $h \leq \lz/2$
in the fracture zone. All simulations were run with 64-bit floating-point
precision. Sections~\ref{sec:dsent} to \ref{sec:perforated}
present the single-solver physics validation against published
benchmarks. Section~\ref{sec:three_way} provides a qualitative
shared-mesh comparison against Akantu and an explicit-dynamic FEniCS
port for matching configurations.

\subsection{Dynamic single-edge notch tension}
\label{sec:dsent}

\paragraph{Problem setup.}
The domain is a $40 \times 40$~mm square glass plate with a horizontal
edge notch of length $a = 20$~mm at mid-height, subjected to a
smooth-step Dirichlet ramp of $\pm 0.001$~mm applied to the top and
bottom surfaces over the first $20~\mu$s and held constant
thereafter. Material parameters and the AT2/spectral/plane-strain
formulation follow Borden et al.~\cite{borden2012phase}:
$E = 32$~GPa,
$\nu = 0.2$, $\rho = 2450$~kg/m$^3$, $\GC = 3$~J/m$^2$, plane
strain, with the AT2 model and the Miehe spectral energy
split~\cite{miehe2010thermodynamically}.

\begin{figure}[htbp]
  \centering
  \includegraphics[width=\textwidth]{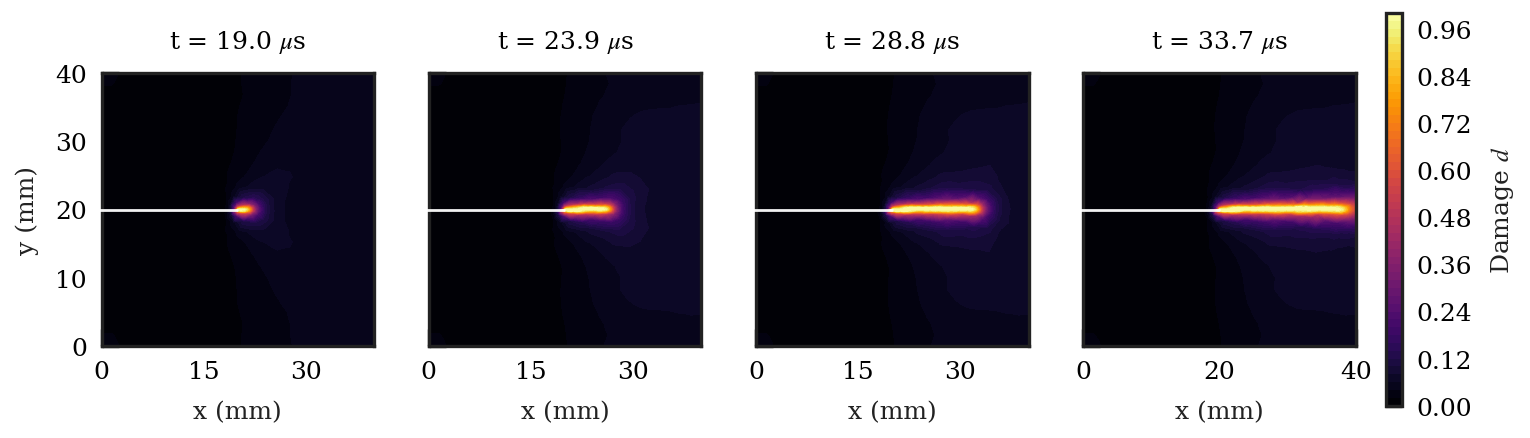}
  \caption{Dynamic SENT: damage field at $t \approx 19.0$, $23.9$,
  $28.8$, $33.7~\mu$s, spanning initiation through the propagating
  straight crack as it reaches the right boundary
  ($h_{\mathrm{crack}} = \lz/2 = 0.25$~mm, $1{,}091$ nodes).}
  \label{fig:dsent}
\end{figure}

\begin{figure}[htbp]
  \centering
  \begin{subfigure}[b]{0.48\linewidth}
    \includegraphics[width=\linewidth]{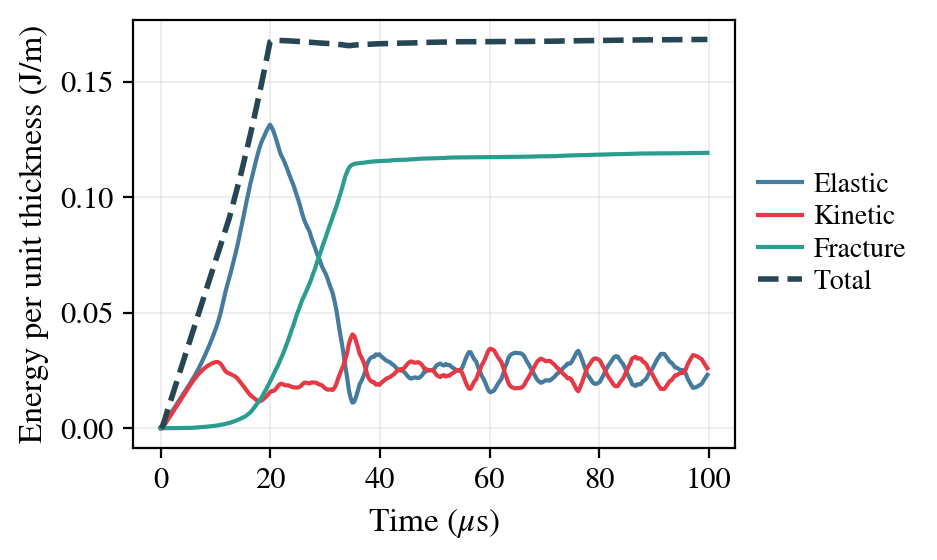}
    \caption{Energy components}
  \end{subfigure}
  \hfill
  \begin{subfigure}[b]{0.48\linewidth}
    \includegraphics[width=\linewidth]{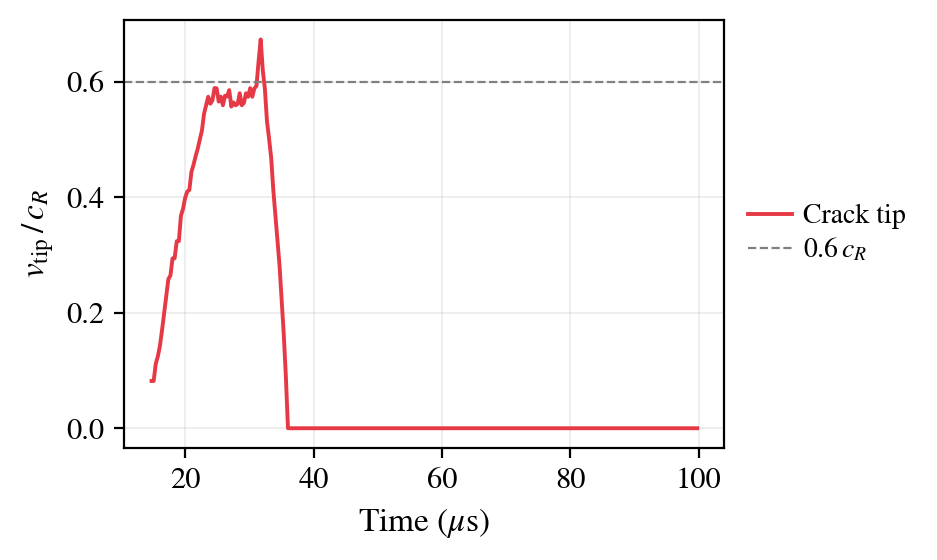}
    \caption{Crack-tip velocity}
  \end{subfigure}
  \caption{Dynamic SENT: energy balance and crack tip velocity
  ($h_{\mathrm{crack}} = \lz/2 = 0.25$~mm, $1{,}091$ nodes).}
  \label{fig:dsent_energy}
\end{figure}

\paragraph{Results and discussion.}
The straight mode-I crack from the notch uses $\lz = 0.5$~mm, with the damage field shown at four snapshots in Fig.~\ref{fig:dsent}.
The crack reaches the right boundary at
$t \approx 33.7\,\mu$s. Over the central propagation interval
($x \approx 25$ to $35$~mm), the smoothed crack-tip speed is
approximately $0.56\,c_R$. The final approach to the right boundary
produces a short tracker spike and is not used as a
steady-propagation measure (Fig.~\ref{fig:dsent_energy}).

Figure~\ref{fig:dsent_energy} shows the energy components and crack-tip
velocity for the dynamic SENT test. The finite-element calculation is
two-dimensional and integrates the volumetric energy density over the
element area $dA$ only. No explicit thickness multiplier is applied, so
the plotted energies are per unit out-of-plane thickness. Equivalently,
integrating J/m$^3$ over a two-dimensional domain gives J/m. A physical
thickness $h_t$ would multiply these values by $h_t$, so a $1$~mm-thick
specimen would have the same numerical value in mJ.

In the plots,
$E_{\mathrm{tot}} = E_{\mathrm{el}} + E_{\mathrm{frac}} +
E_{\mathrm{kin}}$. The elastic energy per unit thickness peaks at
$\sim 0.13$~J/m around $t \approx 20~\mu$s and then decays as the crack
propagates. The crack-surface energy increases monotonically to
$\sim 0.12$~J/m, which is consistent with a straight $40$~mm crack at
$\GC = 3$~J/m$^2$. The energy components therefore show the expected
transfer from stored elastic energy into crack-surface dissipation and
kinetic energy during propagation. The crack-tip velocity rises rapidly
after initiation and oscillates around $0.55$ to $0.60\,c_R$ during
stable propagation. The boundary arrival at
$t \approx 33\,\mu\text{s}$ produces a brief velocity spike in the
crack-tip tracker as the crack zone approaches the free edge.

\subsection{Kalthoff--Winkler impact test}
\label{sec:kalthoff}

\begin{figure}[htbp]
  \centering
  \includegraphics[width=\linewidth]{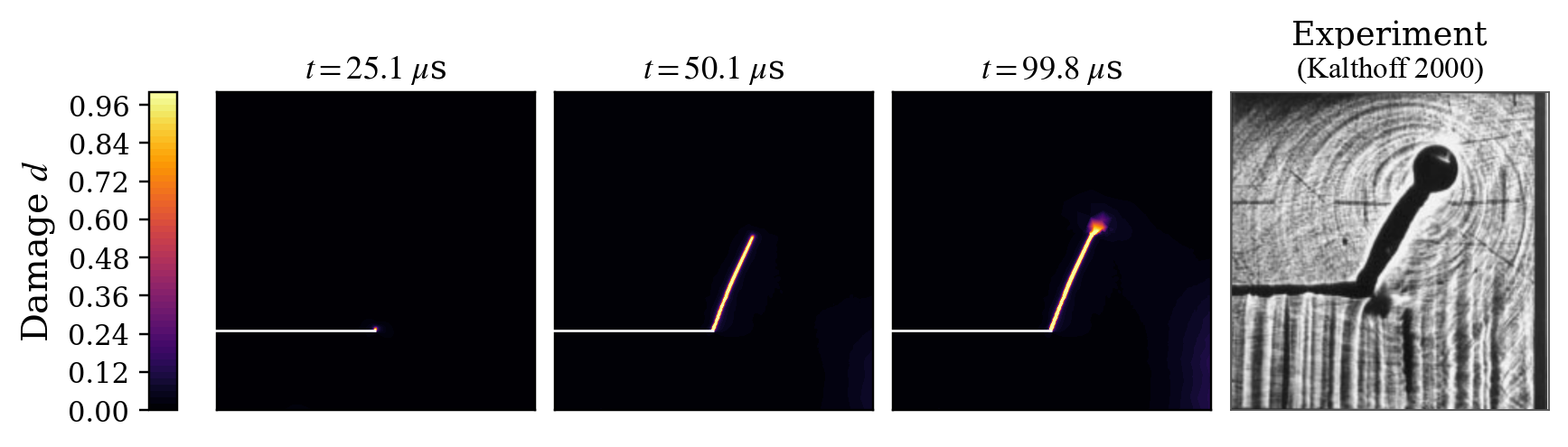}
  \caption{Kalthoff--Winkler benchmark: simulation vs experiment.
  Left three panels: present-solver damage field $d$ at
  $t \approx 25$, $50$, and $100~\mu$s on mesh~3
  ($h = 0.25$~mm, $35{,}487$ nodes); the notch at $y = 25$~mm is
  visible as the horizontal white line, the symmetry boundary is at
  $y = 0$. Right panel: experimental shadow photograph from a
  low-rate LECEI test with high-strength steel~\cite{kalthoff2000modes};
  the simulated and
  experimental cracks kink off the notch at the same
  ${\sim}70^\circ$ angle (consistent with Fig.~13 of Borden et
  al.~\cite{borden2012phase}).
  Experimental timing data (initiation, crack-tip velocity history)
  are not reported in the experimental reference~\cite{kalthoff2000modes}, so the
  quantitative validation in Table~\ref{tab:kalthoff} is restricted
  to the kink angle.}
  \label{fig:kalthoff_snapshots}
\end{figure}

\paragraph{Problem setup.}
The Kalthoff--Winkler experiment~\cite{kalthoff2000modes} subjects a
pre-notched steel plate to projectile impact, producing a dynamic
shear loading at the notch tips. Kalthoff~\cite{kalthoff2000modes} introduces the
Loading rate Effects on Crack Extension Initiation (LECEI) variant
at intermediate impact rates, which is the dataset whose shadow
photograph is reproduced in Fig.~\ref{fig:kalthoff_snapshots} for
qualitative comparison; the geometry and material below follow
Section~4.3 of Borden et al.~\cite{borden2012phase} for direct
numerical reproduction, with the dynamic phase-field formulation of
Hofacker and Miehe~\cite{hofacker2013phase} used as the numerical
reference setting. The full specimen is $100 \times 200$~mm
with two parallel edge notches of depth $a = 50$~mm at $y = 75$~mm
and $y = 125$~mm and impact on the left edge between them. Exploiting
the horizontal mirror symmetry of geometry and loading about the
mid-plane $y = 100$~mm, only the upper half is modelled: a
$100 \times 100$~mm domain with one notch at $y = 25$~mm, an
imposed $u_y = 0$ symmetry condition on the bottom edge ($y = 0$),
and the impact applied on the left-edge segment below the notch
($y \in [0, 24]$~mm). This halves the degrees of freedom without
loss of information relative to the full-plate setup.

The impact is modelled as a prescribed velocity on the impact segment,
ramped linearly from~$0$ to $v_0 = 16.5$~m/s over $t_r = 1$~$\mu$s
and held constant thereafter. Material: $E = 190$~GPa, $\nu = 0.3$,
$\rho = 8000$~kg/m$^3$, $\GC = 2.213 \times 10^4$~J/m$^2$,
$\lz = 0.195$~mm. Five half-plate meshes with $h$ from 1.0~mm down
to 0.1~mm span a 70$\times$ node-count sweep; simulation time
$T = 100$~$\mu$s.

\paragraph{Results.}
Figure~\ref{fig:kalthoff_snapshots} shows the damage field at three
representative simulation times from the half-plate mesh~3 run
($h = 0.25$~mm, $35{,}487$ nodes). The crack initiates at the notch
tip just before $t \approx 25~\mu$s and propagates into the specimen
at an angle exceeding $65^\circ$ from the horizontal, reproducing
the sequence in Fig.~13 of Borden et al.~\cite{borden2012phase}.

Table~\ref{tab:kalthoff} reports the computed crack angles and
initiation times across the five half-plate meshes. The per-mesh
ranges span $67^\circ$ to $73^\circ$ overall, with mesh~1 at
$70^\circ$ to $73^\circ$ and mesh~5 at $68^\circ$ to $70^\circ$. All
meshes are consistent with the numerical result of Borden et
al.~\cite{borden2012phase} ($>65^\circ$) and the experimental value
($\sim 70^\circ$~\cite{kalthoff2000modes}). Initiation times
converge to $\sim 24~\mu$s as the mesh is refined, in close
agreement with the reported $\sim 25~\mu$s reference value.

\begin{table}[htb]
\centering
\caption{Kalthoff--Winkler half-plate mesh study. Crack angles are
measured from the notch tip to the crack tip at $t = 100~\mu$s.
Initiation time is the first instant at which the crack tip leaves the
notch tip ($d > 0.5$ for $x > a$). The reported angle range reflects
the sensitivity of automated crack-tip extraction to the diffuse
phase-field contour near the notch.}
\label{tab:kalthoff}
\begin{tabular}{@{}lrrrr@{}}
\toprule
Mesh & $h$ (mm) & Nodes & Crack angle & Initiation ($\mu$s) \\
\midrule
1 & 1.00  & 3{,}015    & $70^\circ$ to $73^\circ$ & 27.9 \\
2 & 0.50  & 9{,}735    & $69^\circ$ to $72^\circ$ & 25.6 \\
3 & 0.25  & 35{,}487   & $67^\circ$ to $72^\circ$ & 24.6 \\
4 & 0.15  & 94{,}878   & $68^\circ$ to $71^\circ$ & 24.3 \\
5 & 0.10  & 213{,}045  & $68^\circ$ to $70^\circ$ & 24.0 \\
\midrule
\multicolumn{2}{l}{Borden et al.~\cite{borden2012phase}} & & $> 65^\circ$ & $\sim 25$ \\
\multicolumn{2}{l}{Experiment~\cite{kalthoff2000modes}} & & $\sim 70^\circ$ & n/a \\
\bottomrule
\end{tabular}
\end{table}

\begin{figure}[htbp]
  \centering
  \includegraphics[width=0.48\linewidth]{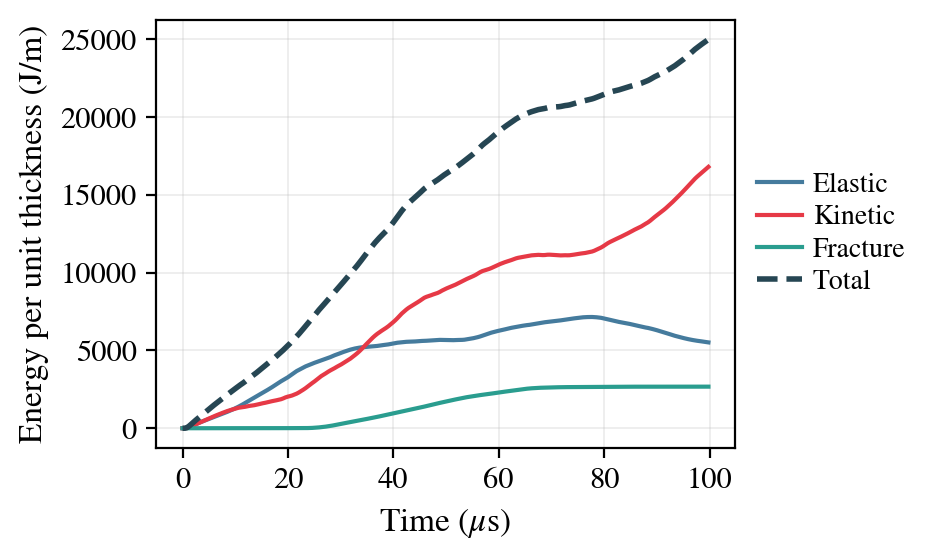}
  \caption{Kalthoff--Winkler half-plate: elastic, kinetic, crack-surface,
  and mechanical energy $E_{\mathrm{tot}}=E_{\mathrm{el}}+E_{\mathrm{frac}}+E_{\mathrm{kin}}$
  per unit out-of-plane thickness vs time
  (mesh~3, $h = 0.25$~mm, $35{,}487$
  nodes).}
  \label{fig:kalthoff_energy}
\end{figure}

Figure~\ref{fig:kalthoff_energy} shows the energy components from the representative
mesh~3 run ($h = 0.25$~mm). The
crack-angle and initiation-time comparison is reported across all five
meshes in Table~\ref{tab:kalthoff}. Energies are again per unit
out-of-plane thickness. The elastic energy
builds up during the initial
stress-wave transit and then partially converts into kinetic and
dissipated crack-surface energy as the crack propagates. The monotonic
growth of the crack-surface term is consistent with irreversible damage
evolution in the phase-field model.

\subsection{Dynamic crack branching}
\label{sec:borden_branching}

\paragraph{Problem setup.}
The dynamic crack-branching benchmark uses a full $100 \times 40$~mm plate with a
horizontal $a = 50$~mm notch at mid-height~\cite{comsol_dynamic_branching64}.
The material parameters are the soda-lime glass values used in
Section~\ref{sec:dsent}:
$E = 32$~GPa, $\nu = 0.2$, $\rho = 2450$~kg/m$^3$,
$\GC = 3$~J/m$^2$, and $\lz = 0.5$~mm. The run uses the AT1 model with
the Amor volumetric--deviatoric split under plane strain, smooth
opposing top/bottom tractions, residual stiffness $\eta = 10^{-7}$, and
central-difference explicit dynamics. The mesh contains 169{,}077 nodes
and 336{,}266 triangular elements.

\begin{figure}[htbp]
  \centering
  \includegraphics[width=\linewidth]{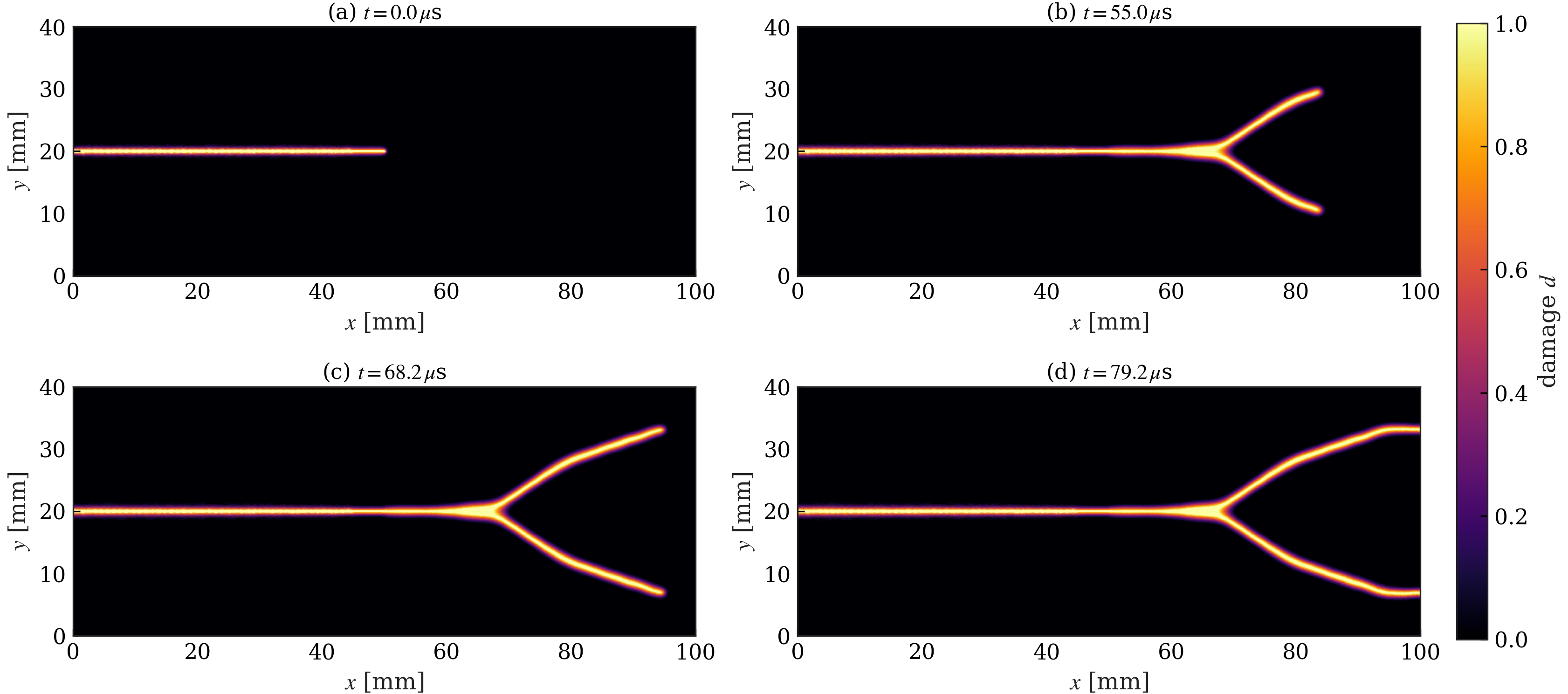}
  \caption{Dynamic crack branching: damage snapshots from the
  full-plate AT1/Amor/plane-strain run. The initially straight notch
  evolves into a symmetric Y-shaped branch.}
  \label{fig:borden_branching_damage}
\end{figure}

\paragraph{Results.}
Figure~\ref{fig:borden_branching_damage} shows a representative
single-run damage sequence. The crack starts from the notch, then
splits into a clear Y-shaped branching pattern while preserving the
same matrix-free explicit mechanics and bound-constrained damage solve
used in the other dynamic benchmarks. Branching onset is detected at
$79.2\,\mu$s, compared with the $68.2$--$70.1\,\mu$s replication
range reported by Ren et al.~\cite{ren2019explicit}. The elastic-energy
peak is $0.258$~J at the COMSOL one-metre-thickness convention
(equivalently $0.258$~J/m per unit thickness), close to the
$0.26$--$0.28$~J reference band used for the same convention. This
example is therefore reported as a dynamic branching morphology and
timing comparison.

\subsection{Crack propagation through perforated plates}
\label{sec:perforated}

\paragraph{Problem setup.}
The PMMA material and plate geometry follow Bleyer et
al.~\cite{bleyer2017dynamic}: $E = 3.09$~GPa, $\nu = 0.35$,
$\rho = 1180$~kg/m$^3$, $\GC = 300$~J/m$^2$, $\lz = 0.1$~mm, and a
$32 \times 16$~mm pre-strained plate with a horizontal edge notch of
length $a = 4$~mm at mid-height. Arrays of circular holes
(diameter $D = 0.4$~mm) are placed along or near the mid-plane of the
specimen. This benchmark is based on the perforated-plate studies in
Sections~4.2 and 4.3 of Bleyer et al.~\cite{bleyer2017dynamic}, which
investigate how holes interact with a propagating crack. The present
work uses representative hole arrays chosen to reproduce the reported
geometric spacings, offsets, loading levels, and qualitative interaction
mechanisms. Two categories of configurations are studied:

\begin{enumerate}[nosep]
  \item \textbf{Holes on the mid-plane} (constraining the crack path):
    two representative arrays are placed on the mid-plane ahead of the
    crack tip. The first uses closely spaced holes
    ($S = 0.9$~mm) and the second uses wider spacing
    ($S = 2.55$~mm), following the constrained-propagation mechanisms
    discussed around Figures~13 to 17 of Bleyer et
    al.~\cite{bleyer2017dynamic}.
  \item \textbf{Distant heterogeneities} (perturbing the crack path):
    a single hole offset 0.6~mm from the mid-plane is placed at
    1~mm or 6~mm from the pre-notch tip. Additional representative
    arrays use holes offset 0.5 or 0.6~mm from the mid-plane with
    $S = 1.95$~mm spacing. These cases follow the distant-heterogeneity
    mechanisms discussed in Section~4.3 and Figures~18 to 19 of
    Bleyer et al.~\cite{bleyer2017dynamic}.
\end{enumerate}

Table~\ref{tab:perforated} summarises the eight simulated
configurations. The hole counts identify the present meshes and are
reported for reproducibility.

\begin{table}[htb]
\centering
\caption{Perforated plate configurations simulated in the present work.
The hole counts define the present meshes used to reproduce the
crack-hole interaction mechanisms discussed by Bleyer et
al.~\cite{bleyer2017dynamic}. All cases use PMMA, AT1, the Amor split,
plane stress, and $D = 0.4$~mm holes.}
\label{tab:perforated}
\small
\begin{tabular}{@{}llrrrl@{}}
\toprule
Config & Present mesh & Offset (mm) & $\Delta U$ (mm) & Nodes & Comparison \\
\midrule
B4a & 30, on mid-plane  & 0   & 0.05 & 195k & Bleyer et al.~\cite{bleyer2017dynamic}, Figure~14 \\
B4b & 10, on mid-plane  & 0   & 0.05 & 183k & Bleyer et al.~\cite{bleyer2017dynamic}, Figure~17 \\
B4c & 1, near notch     & 0.6 & 0.04 & 50k  & Bleyer et al.~\cite{bleyer2017dynamic}, Figure~18a \\
B4d & 1, far from notch & 0.6 & 0.04 & 53k  & Bleyer et al.~\cite{bleyer2017dynamic}, Figure~18b \\
B4e & 15, off-centre    & 0.5 & 0.04 & 214k & Bleyer et al.~\cite{bleyer2017dynamic}, Figure~19a \\
B4f & 15, off-centre    & 0.5 & 0.05 & 214k & Bleyer et al.~\cite{bleyer2017dynamic}, Figure~19b \\
B4g & 15, off-centre    & 0.6 & 0.04 & 214k & Bleyer et al.~\cite{bleyer2017dynamic}, Figure~19c \\
B4h & 15, off-centre    & 0.6 & 0.05 & 214k & Bleyer et al.~\cite{bleyer2017dynamic}, Figure~19d \\
\bottomrule
\end{tabular}
\end{table}

\begin{figure}[htbp]
  \centering
  \begin{subfigure}[b]{\linewidth}
    \includegraphics[width=\linewidth]{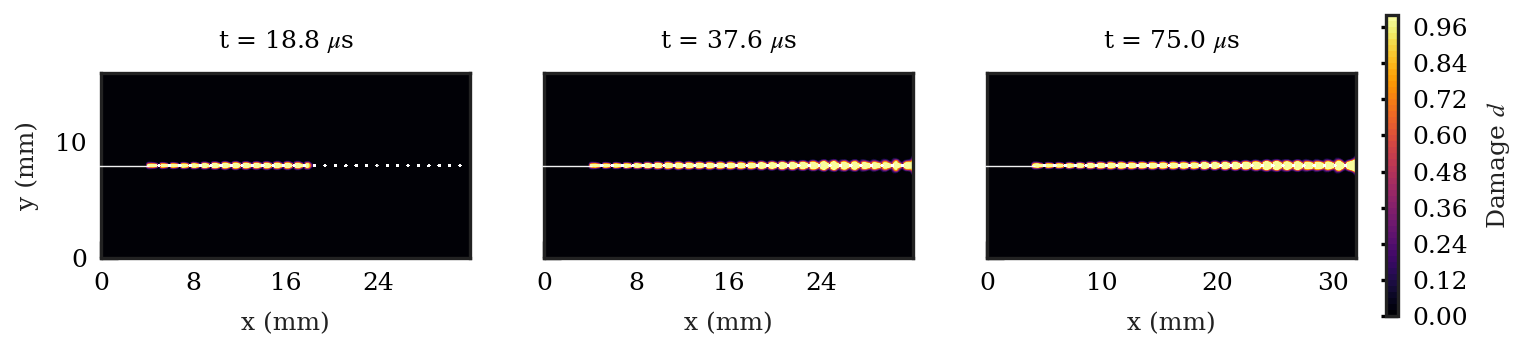}
    \caption{B4a, 30 holes ($S = 0.9$~mm), $\Delta U = 0.05$~mm}
  \end{subfigure}\\[0.6em]
  \begin{subfigure}[b]{\linewidth}
    \includegraphics[width=\linewidth]{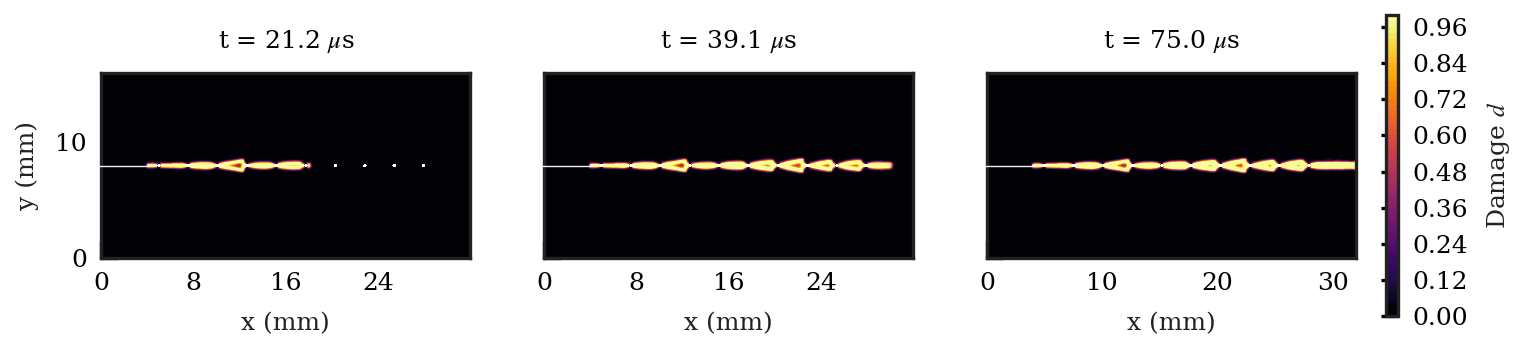}
    \caption{B4b, 10 holes ($S = 2.55$~mm), $\Delta U = 0.05$~mm}
  \end{subfigure}
  \caption{Constrained crack propagation through perforated plates
  with holes on the mid-plane.}
  \label{fig:perf_constrained}
\end{figure}

\paragraph{Constrained propagation (holes on mid-plane).}
Figure~\ref{fig:perf_constrained} shows the damage field for the 30-hole
configuration B4a and the 10-hole configuration B4b. When 30
closely-spaced holes are present, the
crack is attracted to each hole in succession, producing a straight
propagation path along the weakened interface. Macroscopic branching is
not sustained, and small branching attempts are visible at the
end of the plate, corresponding to events D and E in Figure~14 of
Bleyer et al.~\cite{bleyer2017dynamic}. The 10-hole
configuration, with wider hole spacing, shows the same hole-attraction
mechanism with wider intervals between interaction events. Bleyer et
al.~\cite{bleyer2017dynamic} do not report a damage-snapshot sequence for
this 10-hole case. They instead report the instantaneous crack-tip
velocity in their Figure~17. Our damage field is qualitatively
consistent with the interaction mechanism inferred from that evidence,
because the crack advances from hole to hole and produces the
same sequence of accelerations before each hole and pauses at the hole
boundary before renucleating on the opposite side.

\begin{figure}[htbp]
  \centering
  \begin{subfigure}[b]{0.48\linewidth}
    \includegraphics[width=\linewidth]{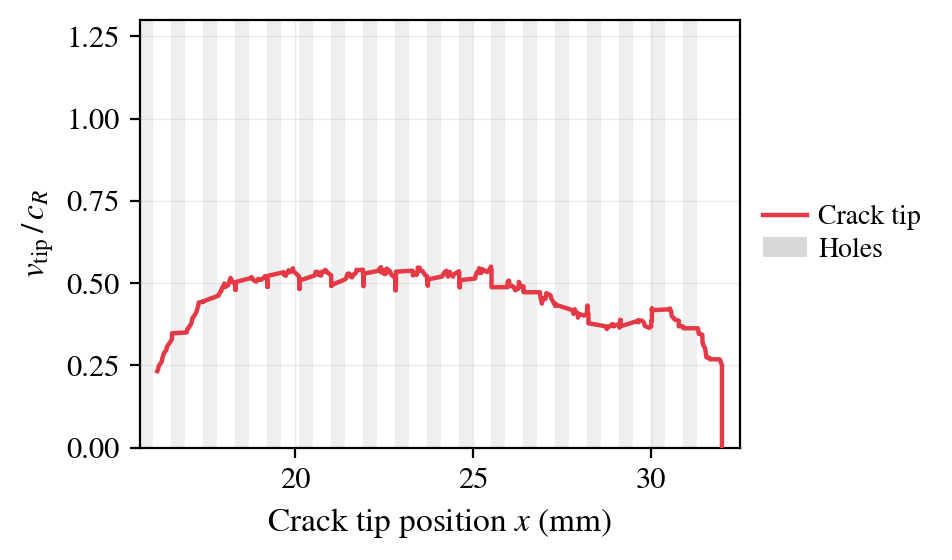}
    \caption{Crack-tip velocity with hole positions}
  \end{subfigure}\hfill
  \begin{subfigure}[b]{0.48\linewidth}
    \includegraphics[width=\linewidth]{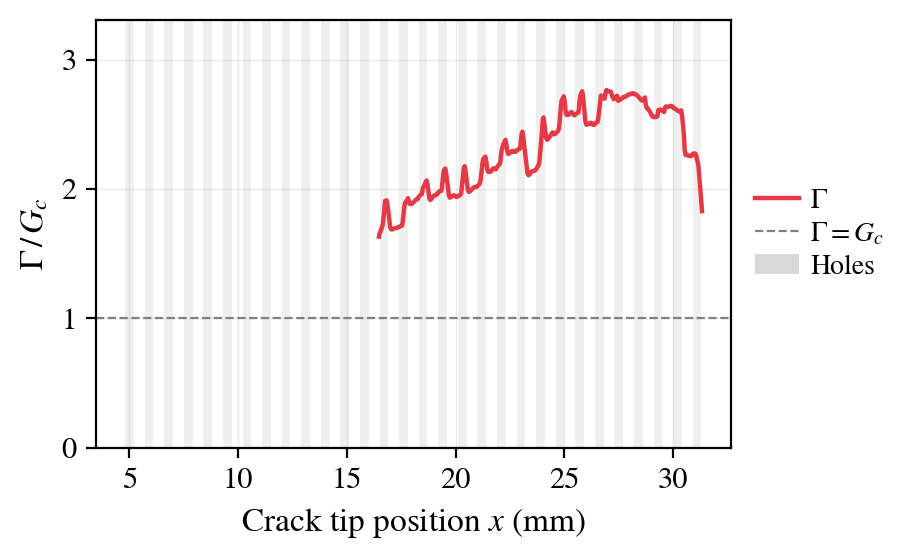}
    \caption{Dissipation per unit crack extension}
  \end{subfigure}\\[0.6em]
  \begin{subfigure}[b]{0.70\linewidth}
    \includegraphics[width=\linewidth]{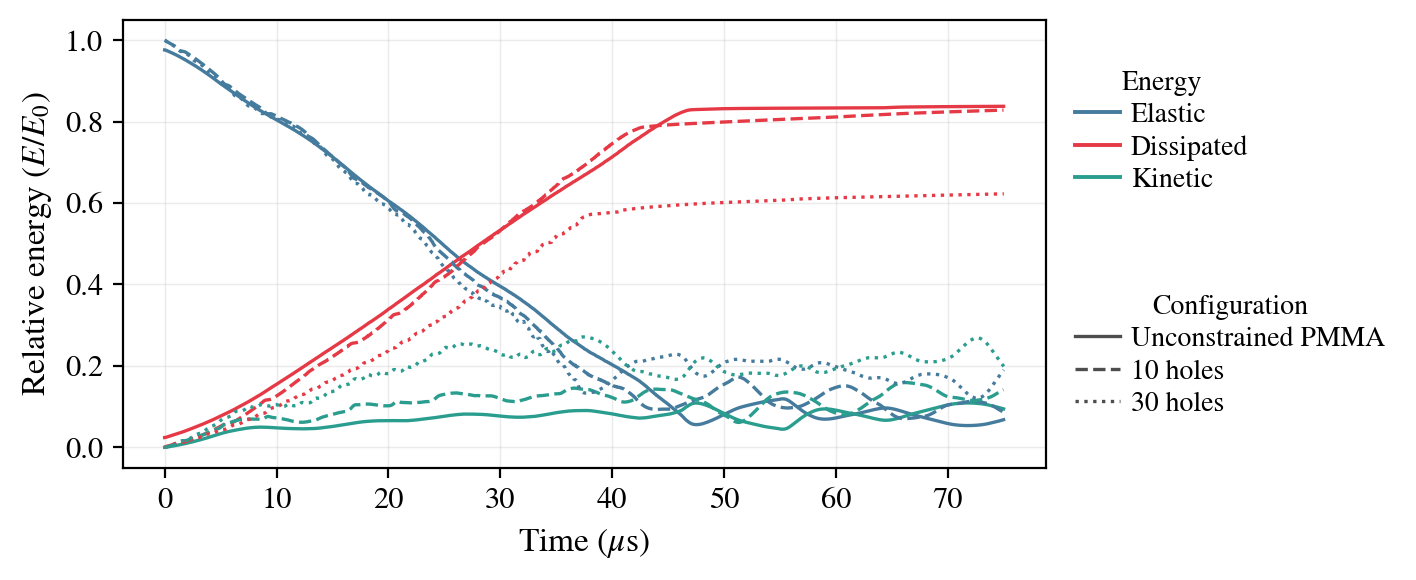}
    \caption{Normalised energy histories ($E/E_0$) with the unconstrained
    baseline (solid), 10 holes (dashed), 30 holes (dotted)}
  \end{subfigure}
  \caption{30-hole constrained perforated-plate results
  ($\Delta U = 0.05$~mm). (a)~Crack-tip velocity along the hole row;
	  (b)~dissipated crack-surface energy per unit crack extension $\Gamma/\GC$ vs crack-tip
  position; (c)~normalised elastic (blue), dissipated (red) and
  kinetic (green) energies for the unconstrained PMMA baseline, 10-hole,
  and 30-hole configurations. See body text for comparison against
  Figures~15 to 17 of Bleyer et al.~\cite{bleyer2017dynamic}.}
  \label{fig:perf_30holes_diag}
\end{figure}

Figure~\ref{fig:perf_30holes_diag} summarises the 30-hole constrained
benchmark with $\Delta U = 0.05$~mm. Panel~(a) plots the crack-tip
velocity with hole positions overlaid as grey bands. The smoothed
envelope peaks near $0.55\,c_R$, while the underlying per-step signal
retains oscillations near each hole interaction. Panel~(b) shows the
dissipated crack-surface energy per unit crack extension, $\Gamma/\GC$,
which rises sharply as the crack tip renucleates beyond each hole.
Panel~(c) compares normalised energy histories for the unconstrained
PMMA baseline and the 10-hole and 30-hole constrained configurations.
The elastic-energy decay is similar, but the 30-hole case accumulates
less crack-surface dissipation, about $0.60\,E_0$ at $t=75\,\mu$s,
than the 10-hole and unconstrained cases, which are both about
$0.85\,E_0$. The difference is transferred mainly into kinetic energy.
This partition is consistent with Figure~16 of
Bleyer et al.~\cite{bleyer2017dynamic}. The 10-hole case shows the same
qualitative trend, but the departure from the unconstrained baseline is
clearest for the denser 30-hole spacing.

\begin{figure}[htbp]
  \centering
  \begin{subfigure}[b]{\linewidth}
    \includegraphics[width=\linewidth]{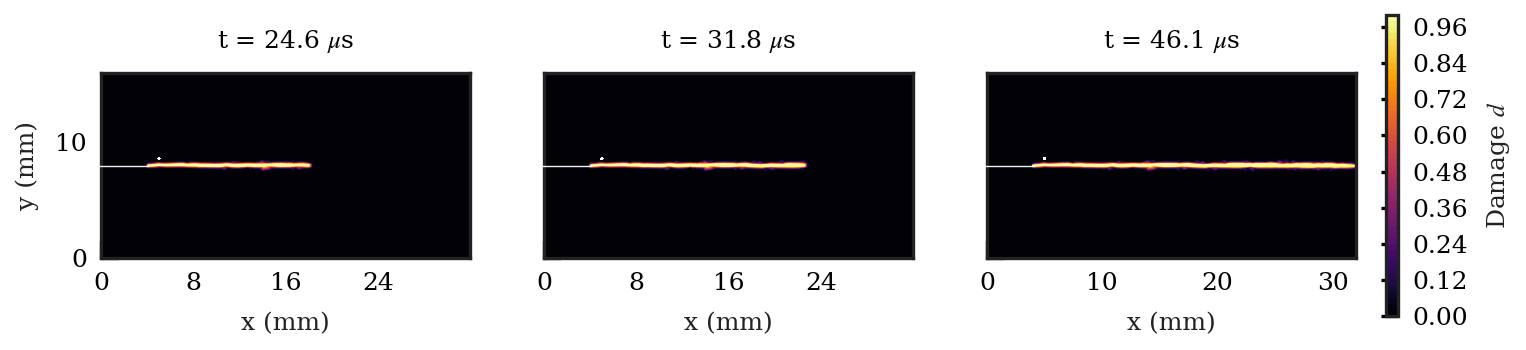}
    \caption{B4c, hole at 1~mm from notch tip (near)}
  \end{subfigure}\\[0.6em]
  \begin{subfigure}[b]{\linewidth}
    \includegraphics[width=\linewidth]{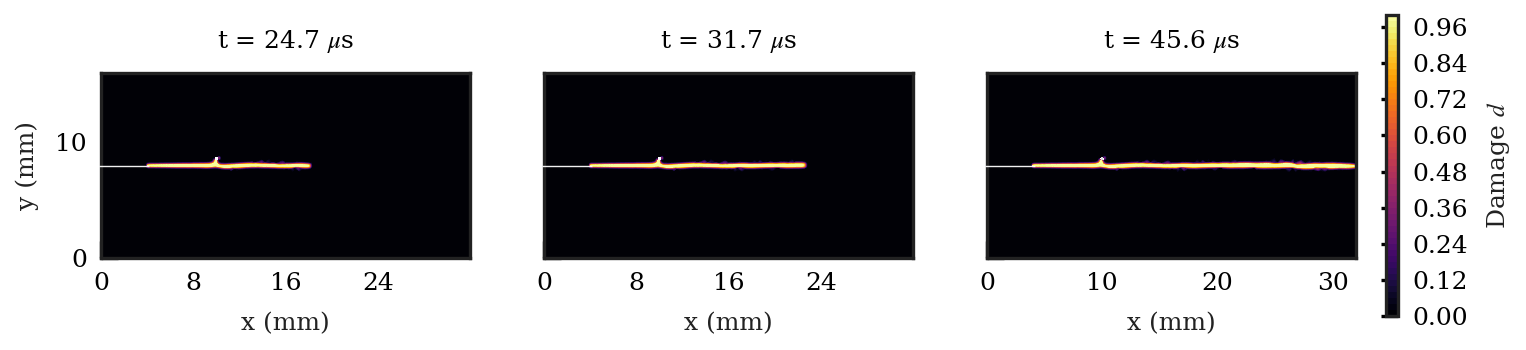}
    \caption{B4d, hole at 6~mm from notch tip (far)}
  \end{subfigure}
  \caption{Single distant hole crack-hole interaction for two hole
  positions ($\Delta U = 0.04$~mm, offset 0.6~mm).}
  \label{fig:perf_1hole}
\end{figure}

\paragraph{Single distant hole.}
Figure~\ref{fig:perf_1hole} shows the crack path for a single hole offset
0.6~mm from the mid-plane, placed at 1~mm (near) and 6~mm (far) from the
pre-notch tip ($\Delta U = 0.04$~mm). For the near hole, the crack passes
close to the hole at low velocity and is only slightly deflected, continuing
its straight path afterward. For the far hole, the crack arrives at
higher velocity and interacts more strongly. A microbranch appears near
the hole, and the crack path is perturbed. The faster crack interacts more
strongly with the distant defect, consistent with Figure~18 of
Bleyer et al.~\cite{bleyer2017dynamic}. This behaviour is explained by the
velocity-toughening mechanism, since a wider damage band at higher
velocities increases the interaction distance.

\begin{figure}[htbp]
  \centering
  \begin{subfigure}[b]{\linewidth}
    \includegraphics[width=\linewidth]{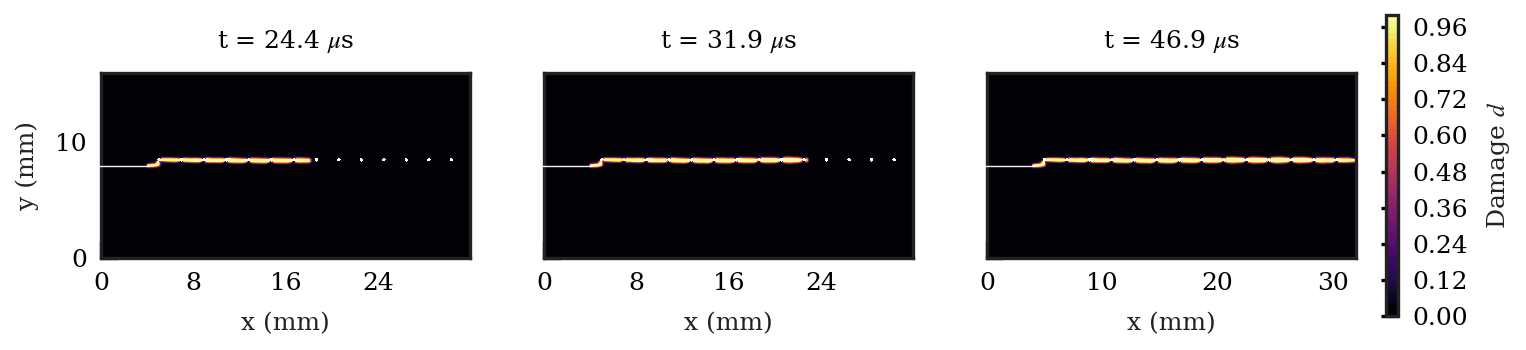}
    \caption{B4e, offset 0.5~mm, $\Delta U = 0.04$~mm}
  \end{subfigure}\\[0.4em]
  \begin{subfigure}[b]{\linewidth}
    \includegraphics[width=\linewidth]{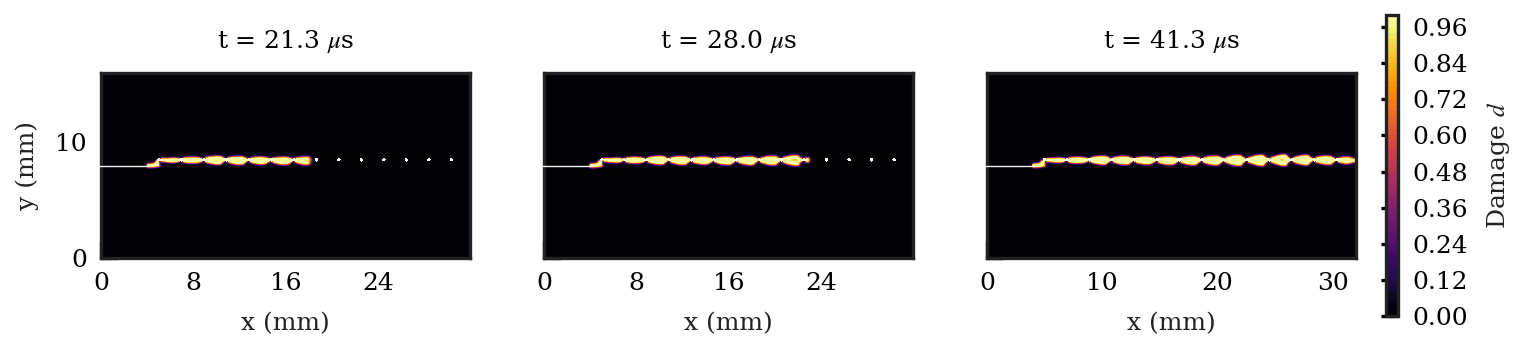}
    \caption{B4f, offset 0.5~mm, $\Delta U = 0.05$~mm}
  \end{subfigure}\\[0.4em]
  \begin{subfigure}[b]{\linewidth}
    \includegraphics[width=\linewidth]{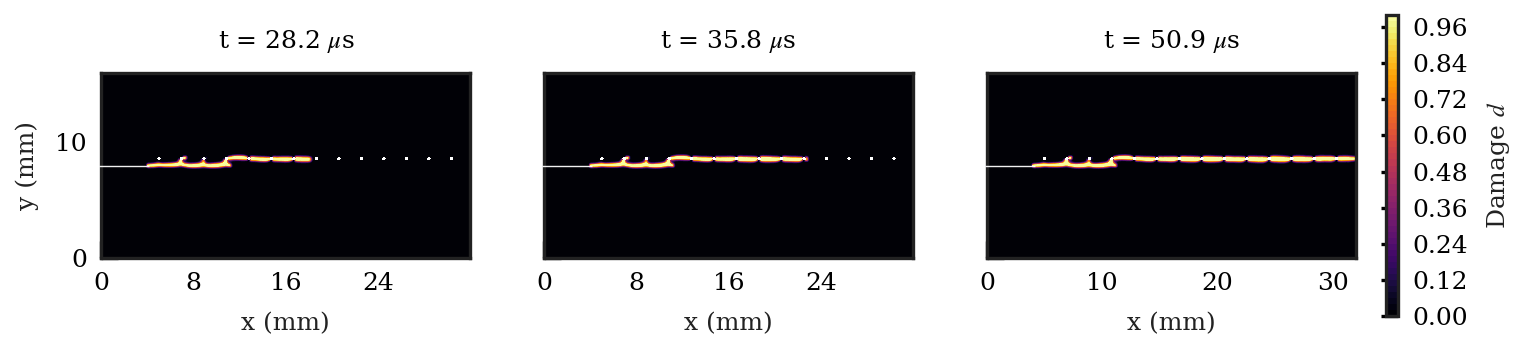}
    \caption{B4g, offset 0.6~mm, $\Delta U = 0.04$~mm}
  \end{subfigure}\\[0.4em]
  \begin{subfigure}[b]{\linewidth}
    \includegraphics[width=\linewidth]{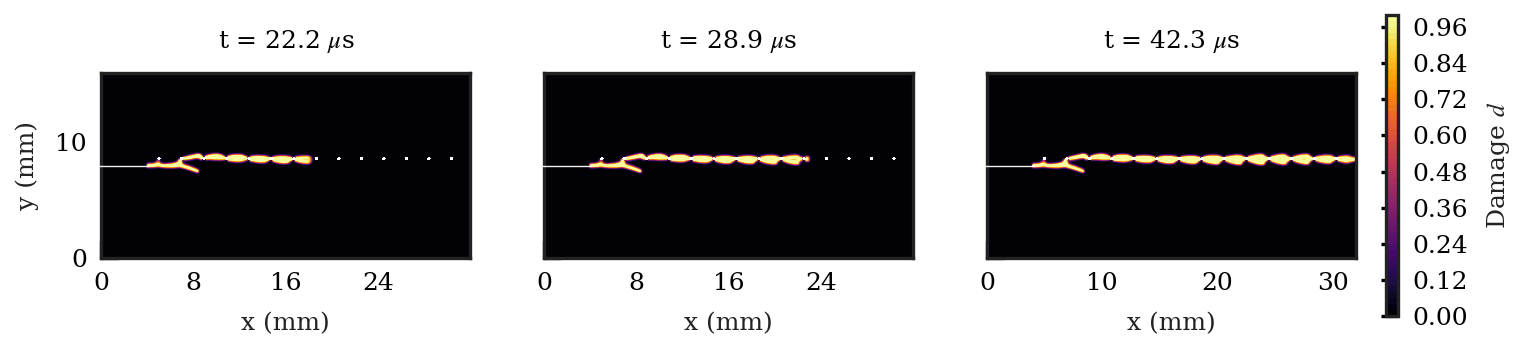}
    \caption{B4h, offset 0.6~mm, $\Delta U = 0.05$~mm}
  \end{subfigure}
  \caption{Crack propagation through 15 distant holes at two offsets
  and two loading levels.}
  \label{fig:perf_15holes}
\end{figure}

\paragraph{Fifteen distant holes.}
The most complex configurations involve 15 holes offset from the mid-plane,
investigated at two offset distances (0.5 and 0.6~mm) and two loading
levels ($\Delta U = 0.04$ and 0.05~mm).
Figure~\ref{fig:perf_15holes} shows the damage fields for all four
combinations. The results demonstrate a strong dependence on the offset
distance:
\begin{itemize}[nosep]
  \item At 0.5~mm offset, the crack localises in the weakened plane
    defined by the holes and propagates along it, similar to the
    constrained mid-plane configurations. This corresponds to
    Figure~19(a,b) of Bleyer et al.~\cite{bleyer2017dynamic}.
  \item At 0.6~mm offset, the holes are sufficiently distant that the
    crack continues straight, with small microbranches emerging toward
    the holes. At higher loading ($\Delta U = 0.05$~mm), these
    microbranches become more pronounced and tend to deflect the main
    crack. This corresponds to Figure~19(c,d) of
    Bleyer et al.~\cite{bleyer2017dynamic}.
\end{itemize}

\subsection{Quasi-static benchmark checks}
\label{sec:qs_benchmarks}

The implementation also includes a staggered quasi-static mode for
rate-independent fracture. Two quasi-static checks are included here as
secondary validation. They are single-edge-notched tension and a
notched-holed plate benchmark~\cite{comsol_holed_plate64}.

\begin{figure}[t]
  \centering
  \includegraphics[width=\linewidth]{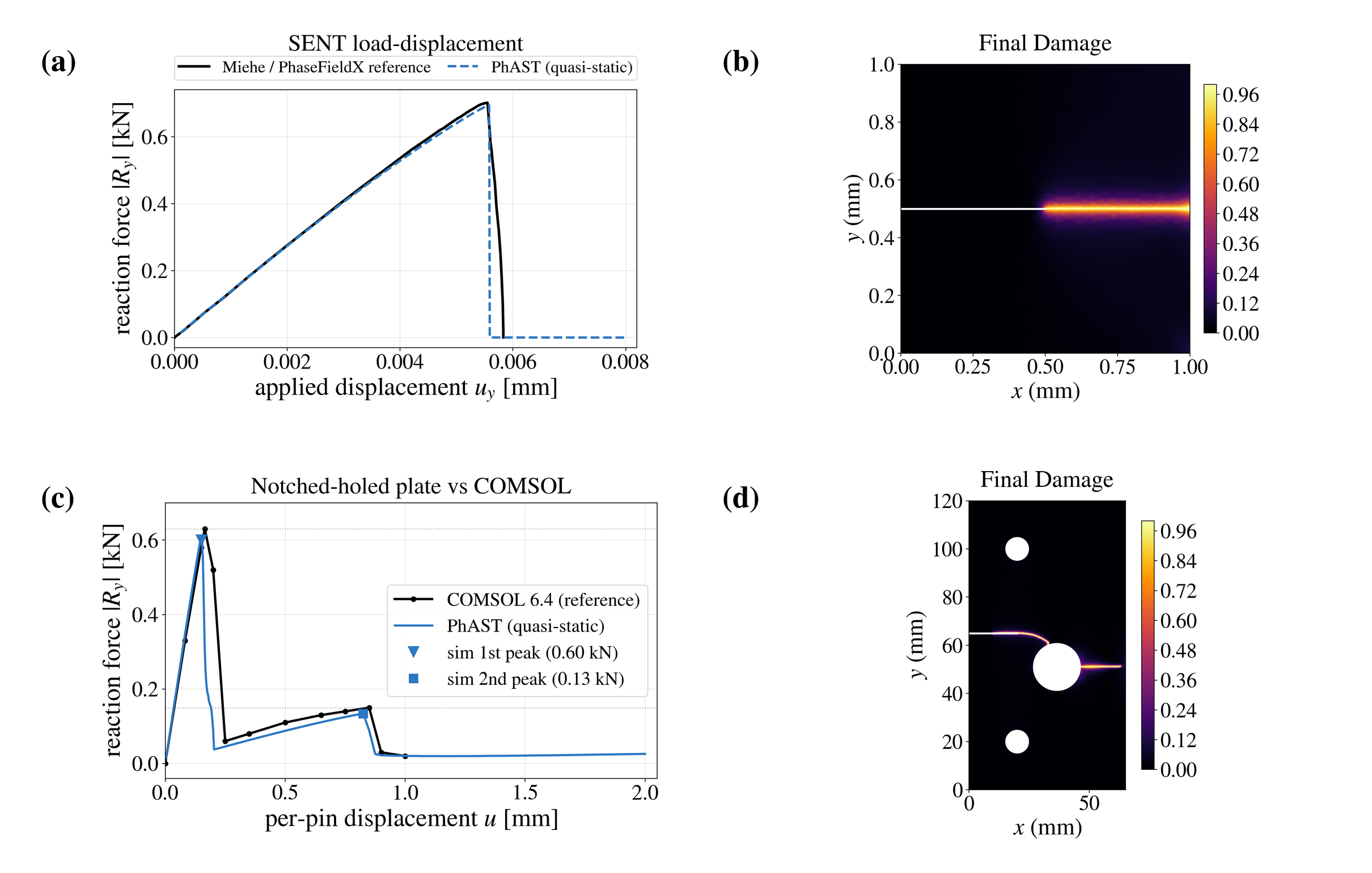}
  \caption{Quasi-static benchmark results. Panels (a,b) show the
  single-edge-notched tension benchmark, compared with the
  PhaseFieldX-style reference with $1.08\%$ peak-load error and
  $1.70\%$ pre-peak $L^2$ error. Panels (c,d) show the
  notched-holed plate benchmark~\cite{comsol_holed_plate64}. The parameter-matched
  comparison run gives $4.68\%$ first-peak load error, $9.09\%$ first-peak displacement error,
  and $10.51\%$ second-peak load error. The crack path bends from the notch toward
  the hole, matching the reference morphology (panel d).}
  \label{fig:qs_benchmarks}
\end{figure}

Figure~\ref{fig:qs_benchmarks} summarises the quasi-static validation
checks. Panels (a,c) compare the load--displacement curves. Panels
(b,d) show the final damage fields for the same runs.

In the SENT case, the PhAST curve follows the PhaseFieldX-style
reference up to peak load, and the damage field shows the expected
straight horizontal crack extension. In the notched-holed plate, the
parameter-matched run gives a first-peak load error of $4.68\%$, a
first-peak displacement error of $9.09\%$, and a second-peak load error
of $10.51\%$. The second peak is more sensitive because it occurs after
the crack has turned toward the large hole. At that stage the response
depends on staggered tolerances, the crack-width-to-mesh ratio, and the
details of the reference solution strategy. The important result is that
the solver reproduces both the global load peaks and the curved
mixed-mode crack path from the notch toward the large hole.

\subsection{Cross-code comparison}
\label{sec:three_way}

This subsection uses independent implementations to assess whether the
reported crack morphologies are robust to the solver implementation.
The comparison is intentionally qualitative and uses matched benchmark
definitions for each solver pair.

The SENT cross-code case uses a smaller shared benchmark mesh than the
main validation run, with $9{,}031$ nodes, $17{,}659$ linear triangles,
and $3000$ explicit steps. This case uses AT2 with the
volumetric--deviatoric Amor split, which allows comparison among the
present solver, the FEniCS explicit-dynamic phase-field port, and the
Akantu reference implementation. The Kalthoff--Winkler cross-code case
uses $35{,}654$ nodes, $71{,}114$ linear triangles, and $11{,}775$
explicit steps with the Miehe spectral split. The present solver and
the FEniCS reference are compared on this spectral-split configuration.
The SENT comparison uses a common mesh and configuration record for all
three solvers. The Kalthoff--Winkler comparison uses the matched
spectral-split configuration for the present solver and the FEniCS
reference. In both comparisons, phase-field subcycling is disabled so
that each solver advances the damage field at every explicit step.

\begin{figure}[htbp]
  \centering
  \includegraphics[width=0.72\linewidth]{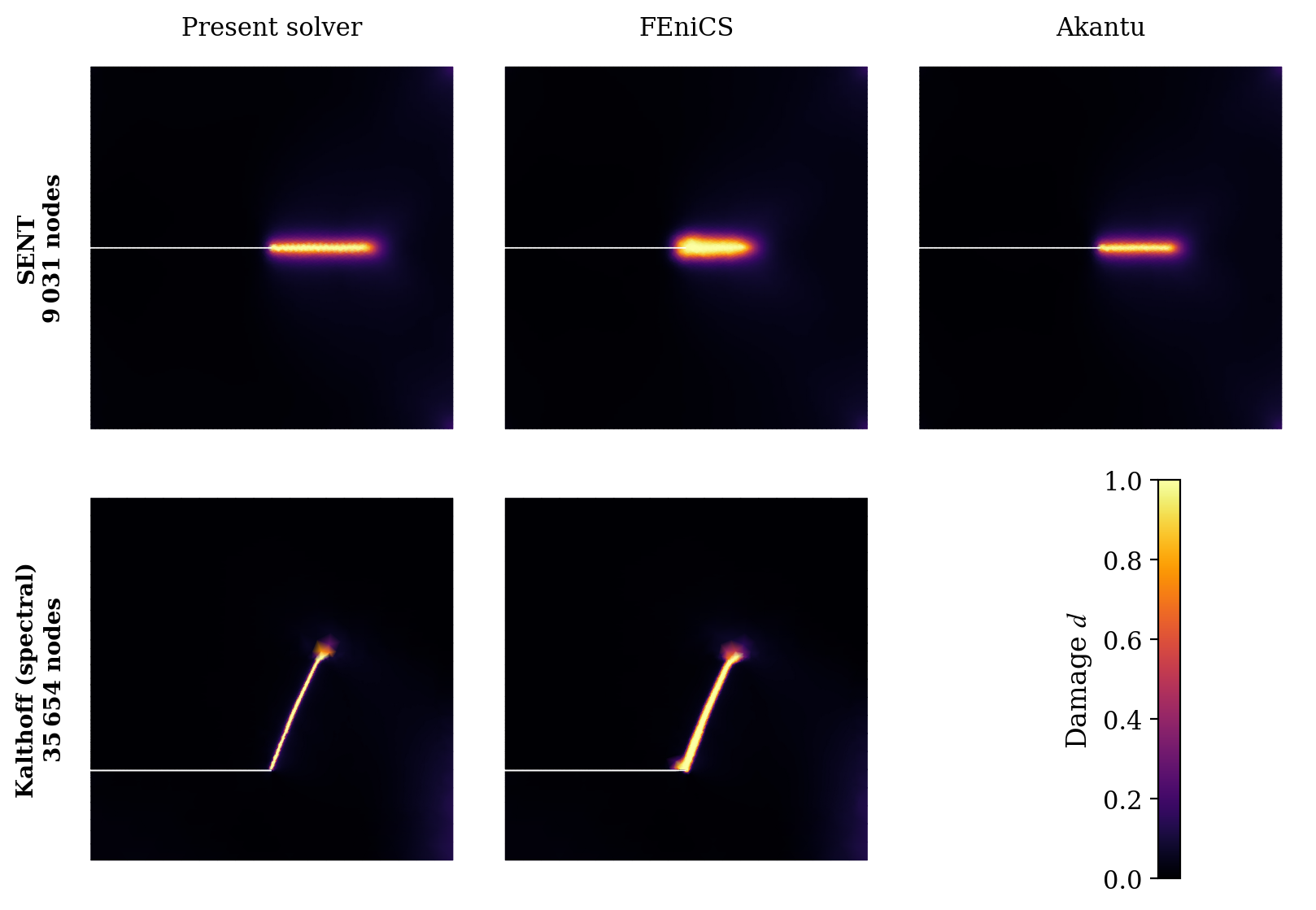}
  \caption{Cross-code solver validation on the two shared benchmarks.
  Final damage field $d$ for (top row) single-edge notched tension
  computed with the present solver (left), an
  open-source FEniCS explicit-dynamic phase-field port (middle),
  and a C++/Message Passing Interface (MPI) Akantu
  reference implementation~\cite{richart2024akantu} (right);
  and (bottom row) Kalthoff--Winkler impact on the half-plate
  geometry with the Miehe spectral split, computed with
  the present solver (left) and the FEniCS port
  (middle). The comparisons use the same benchmark definitions; the
  SENT row uses a shared mesh, while the Kalthoff row compares the
  spectral-split runs from each implementation. Damage is displayed clipped to the physical
  $[0,1]$ range.}
  \label{fig:threeway_montage}
\end{figure}

Figure~\ref{fig:threeway_montage} compares the final damage fields for
the shared SENT and Kalthoff--Winkler configurations. In the SENT case,
the present solver, the FEniCS port, and Akantu all recover a straight
mode-I crack from the edge notch. In the Kalthoff--Winkler case, the
present solver and the FEniCS reference both recover the inclined crack
trajectory associated with the spectral formulation. The role of this
comparison is to check implementation robustness rather than to
introduce a new quantitative error metric.

Table~\ref{tab:benchmark_summary} summarises the validation evidence
from the benchmark section.

\begin{table*}[htb]
\centering
\caption{Benchmark evidence summary. Rows B1 to B4 are explicit
dynamic fracture benchmarks; rows Q1 to Q2 are staggered quasi-static
benchmarks. Setup parameters for all six benchmarks are in
Table~\ref{tab:benchmark_overview}; detailed discussion follows in
Sections~\ref{sec:dsent} to \ref{sec:qs_benchmarks}.}
\label{tab:benchmark_summary}
\small
\setlength{\tabcolsep}{4pt}
\begin{tabular}{@{}
  >{\raggedright\arraybackslash}p{0.18\textwidth}
  >{\raggedright\arraybackslash}p{0.39\textwidth}
  >{\raggedright\arraybackslash}p{0.33\textwidth}@{}}
\toprule
Benchmark & Key result & Reference comparison \\
\midrule
B1: Dynamic SENT & Straight mode~I crack; central-interval smoothed
  speed about $0.56\,c_R$ before boundary-arrival transients &
  Borden et al.~\cite{borden2012phase}; square SENT run with
  $\lz = 0.5$~mm gives the expected straight propagation \\
B2: Kalthoff--Winkler & Angle $67^\circ$ to $73^\circ$ across five meshes
  under per-mesh selection closest to the experimental
  $\sim 70^\circ$ (mesh~5 finest: $68^\circ$ to $70^\circ$) &
  Fig.~13 of Borden et al.~\cite{borden2012phase};
  Hofacker and Miehe~\cite{hofacker2013phase}; Kalthoff~\cite{kalthoff2000modes} \\
B3: Dynamic crack branching & Full-plate run develops a clear
  Y-shaped dynamic branch from the pre-notch; branching onset
  $79.2\,\mu$s and elastic peak $0.258$~J at the COMSOL thickness convention &
  Benchmark setup~\cite{comsol_dynamic_branching64} with material
  values from Borden et al.~\cite{borden2012phase}; dynamic
  branching morphology and reference timing range \\
B4: Perforated plate & Representative 30-/10-hole on-plane locking; 1-hole
  near/far symmetry; 15-hole regimes at 0.5/0.6 mm offset &
  Bleyer et al.~\cite{bleyer2017dynamic}, Figs.~14 to 19 \\
\midrule
Q1: SENT & Peak reaction $0.6936$ vs $0.7012$~kN
  ($1.08\%$); pre-peak $L^2$ error $1.70\%$;
  dissipated-energy error $5.38\%$ &
  Miehe et al.~\cite{miehe2010thermodynamically} and the PhaseFieldX
  example 1711 documentation benchmark~\cite{castillon2025phasefieldx};
  peak response, pre-peak response, dissipated energy, and final damage
  match the reference closely; the unstable snap-back branch is
  shown separately \\
Q2: Notched-holed plate & First peak $0.6005$ vs $0.6300$~kN
  ($4.68\%$ load error) at per-pin displacement
  $u = 0.1500$ vs $0.1650$~mm
  ($9.09\%$ displacement error); second peak $0.1342$ vs
  $0.1500$~kN ($10.51\%$ load error) &
  Holed-plate fracture benchmark~\cite{comsol_holed_plate64};
  the curved crack path from the notch toward the large hole is checked in
  Fig.~\ref{fig:qs_benchmarks} \\
\bottomrule
\end{tabular}
\end{table*}

\section{Differentiability and its applications}
\label{sec:differentiability}

This section explains how the solver can be used as a differentiable
forward model. A simulation is first run forward to produce a fracture
state. A scalar loss or quantity of interest is then computed from that
state. Reverse-mode automatic differentiation gives the derivative of
that scalar quantity with respect to material, geometric, or loading
parameters.

This derivative has a direct inverse-problem interpretation. It
quantifies how a small change in each parameter would change the
simulated observation. This matters because finite differences require a new
forward simulation for each parameter perturbation. For a parameter
vector $\theta \in \mathbb{R}^k$, central finite differences require
$2k$ forward simulations. Reverse-mode automatic differentiation gives
the gradient of one scalar loss with respect to all $k$ parameters using
one forward pass and one backward pass. For phase-field fracture runs
that may take minutes to hours, this difference becomes important as
soon as more than a few parameters are unknown.

The solver supports this workflow because the main operations are
written as PyTorch tensor operations. Element-wise strain evaluation,
scatter-based force assembly, history updates, and matrix-free damage
operators can therefore participate in the autograd graph. The explicit
mechanics update is differentiated directly through these tensor
operations. The iterative damage solve uses an implicit differentiation
rule for the converged linear system, so the backward pass does not need
to store every conjugate-gradient iteration.

The phase-field solve also contains nonsmooth operations. The history
maximum, the damage irreversibility clamp, and the AT1 active-set
projection are treated as piecewise differentiable operations. An
active-set switch occurs when a damage degree of freedom changes between
a free unknown and a constrained one. At such points, the local
linearisation can change abruptly. The gradients reported here should
therefore be read as sensitivities of the realised discrete trajectory.

The demonstrated application is scalar material-parameter inversion for
two different values of fracture toughness. The same differentiable
forward map can also support larger inverse problems, including
geometric localisation of inclusions, spatially varying $\GC(\bs{x})$
recovery, and training workflows that couple the solver to
neural-network models.

\subsection{Material parameter inversion}
\label{sec:inversion}

The inverse problem considered here is the recovery of fracture
toughness from an observed crack pattern. The geometry, boundary
conditions, loading, and all other material parameters are assumed to be
known. The unknown parameter is the scalar fracture toughness $\GC$.

A reference simulation provides the target final damage field
$d^{*}(\bs{x})$. For a trial value of $\GC$, the solver runs the full
forward fracture simulation and produces a predicted final damage field
$d_T(\GC)$. The mismatch between the predicted and target fields is
measured with the nodewise mean-squared error (MSE)
\begin{equation}
  \mathcal{L}(\GC) = \frac{1}{N} \sum_{i=1}^{N}
    \left( d_T^{(i)}(\GC) - d^{*\,(i)} \right)^2,
  \label{eq:inversion_loss}
\end{equation}
where $N$ is the number of nodes. The loss is small when the predicted
crack pattern matches the target field and large when the crack is
misplaced or has the wrong extent.

This loss is used as a simple baseline. It is not a perfect crack-shape
metric. The damage field is close to zero away from the crack and close
to one inside the cracked region. A nodewise error therefore compares
whether the crack occupies the same nodes, rather than comparing only
the overall crack shape. A small lateral shift of an otherwise similar
crack can produce a large loss, because many nodes change from
undamaged to damaged or from damaged to undamaged. This also makes the
objective non-convex. As $\GC$ changes, the crack may initiate at a
slightly different point or stop at a different length, so the loss can
develop local minima associated with misplaced crack tips. The baseline
is sufficient for the scalar inversions reported below. Geometry-aware
losses, such as optimal-transport or Wasserstein distances between
damage fields, can be added in the same autograd pipeline in future
work.

The motivation for using autograd is the cost of the gradient. Without
gradients, sampling-based Bayesian methods~\cite{wu2021parameter,
stanic2026probabilistic}, integrated DIC~\cite{kosin2024parameter},
and field-valued recovery methods~\cite{gao2023inverse} often require
many forward simulations. Finite-difference Jacobians also become
expensive because their cost grows with the number of unknown
parameters. In contrast, the reverse-mode path used here gives the
gradient of $\mathcal{L}$ with respect to all active parameters using
one backward sweep, subject to the active-set caveat described above.

The first test uses the square pre-notched glass SENT variant from
Section~\ref{sec:dsent}. The material parameters are those of
Borden et al.~\cite{borden2012phase}. The mesh contains $1{,}363$
nodes and $2{,}531$ elements, and the forward simulation is run for 200
explicit time steps. The reference simulation uses
$\GC^{*} = 3 \times 10^{-3}$~N/mm. The optimisation starts from
$\GC^{(0)} = 6 \times 10^{-3}$~N/mm, which is twice the true value.

Each loss evaluation reruns the 200-step differentiable forward
simulation. The explicit momentum update is differentiated directly
through PyTorch tensor operations. The damage update is differentiated
through the converged damage equation using the implicit linear-solve
rule. The gradient is therefore obtained through the realised discrete
trajectory without manually perturbing $\GC$.

A central finite-difference calculation is used once at the initial
iterate as an independent gradient check. The autograd derivative
$\partial \mathcal{L}/\partial\log\GC =
2.791115\times 10^{-3}$ agrees with the central finite-difference value
$2.796419\times 10^{-3}$. The relative difference is
$1.90\times 10^{-3}$, which confirms the implemented derivative at the
starting point.

The optimisation uses limited-memory Broyden--Fletcher--Goldfarb--Shanno
(L-BFGS) with a strong-Wolfe line search on
$\log\GC$. Optimising the logarithm keeps the recovered toughness
positive. At each outer iteration, the optimiser reruns the forward
simulation, computes the MSE loss, calls \texttt{backward()}, and then
chooses a stable step length. The method recovers
$\GC = 3.000005\times 10^{-3}$~N/mm after ten optimiser iterations,
with a relative error of $1.72\times10^{-6}$. This final value is not
used as the practical stopping point. The recovery already meets a
$10^{-3}$ relative-error tolerance after three accepted L-BFGS states,
where $\GC = 2.999548\times 10^{-3}$~N/mm. The remaining iterations were
kept to check that the parameter remained on a stable plateau. The full
run uses 19 autograd closure evaluations including line-search probes.

\begin{figure}[htbp]
  \centering
  \includegraphics[width=0.92\textwidth]{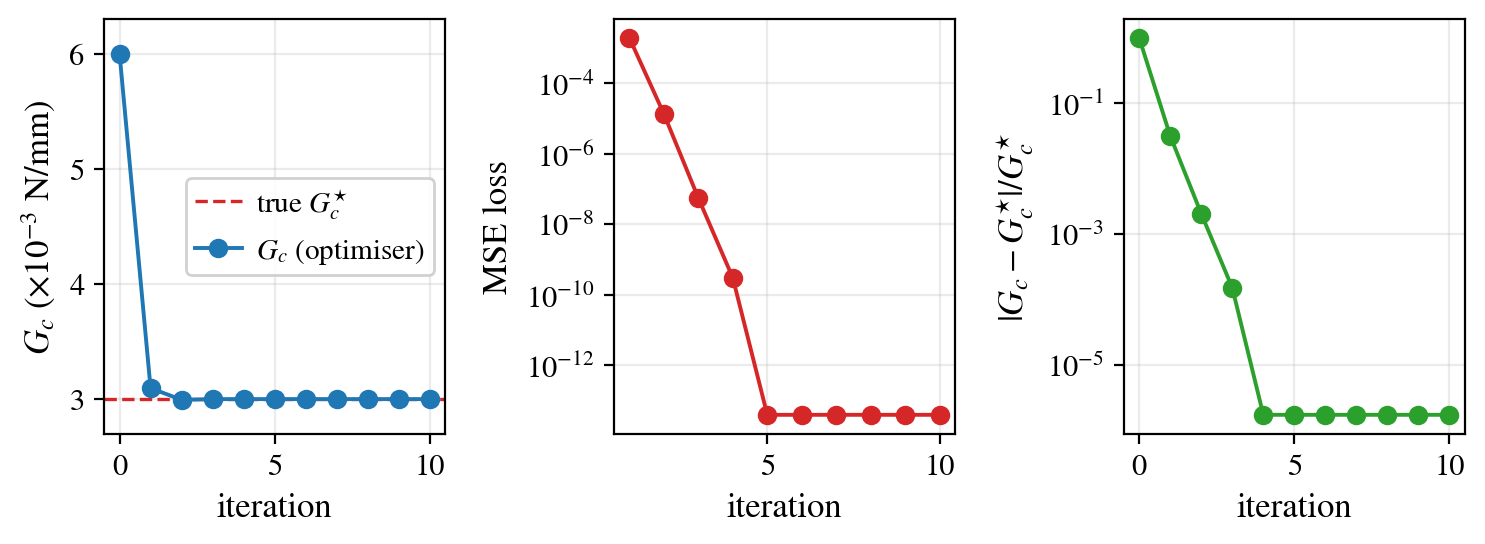}
  \caption{Autograd-driven scalar $\GC$ recovery from an observed final
  damage field. The optimiser starts from
  $\GC^{(0)} = 6\times 10^{-3}$~N/mm ($2\times$ the true value
  $\GC^{*} = 3\times 10^{-3}$~N/mm) and uses PyTorch reverse-mode
  gradients through the differentiable forward solve. Central finite
  differences are used only as an initial gradient check. Iteration~0
  marks the initial guess; the ten subsequent L-BFGS outer iterations
  expose the convergence plateau. The recovery is already below
  $10^{-3}$ relative error after three accepted L-BFGS states.}
  \label{fig:inversion}
\end{figure}

The same recovery is repeated on alumina (Al$_2$O$_3$) to test the
procedure on a second toughness value. The alumina parameters are taken
from Kumar et al.~\cite{kumar2020revisiting}. The reference toughness is
$\GC^{*} = 26.8$~J/m$^2$. The optimisation starts from
$\GC^{(0)} = 54.0$~J/m$^2$, again approximately twice the true value.

Using the same mesh, loss, and L-BFGS driver, the method recovers
$\GC = 26.80008$~J/m$^2$ after ten optimiser iterations, but the
practical recovery criterion is reached earlier. The final relative
error is $3.03\times10^{-6}$. The run uses 30 autograd closure
evaluations including line-search probes. The parameter is already below
$10^{-3}$ relative error after two accepted L-BFGS states. The later
points are retained as a stability check. This second example indicates
that the inversion is driven by the damage-field mismatch rather than by
tuning to one specific value of $\GC$.

\begin{figure}[htbp]
  \centering
  \includegraphics[width=0.92\textwidth]{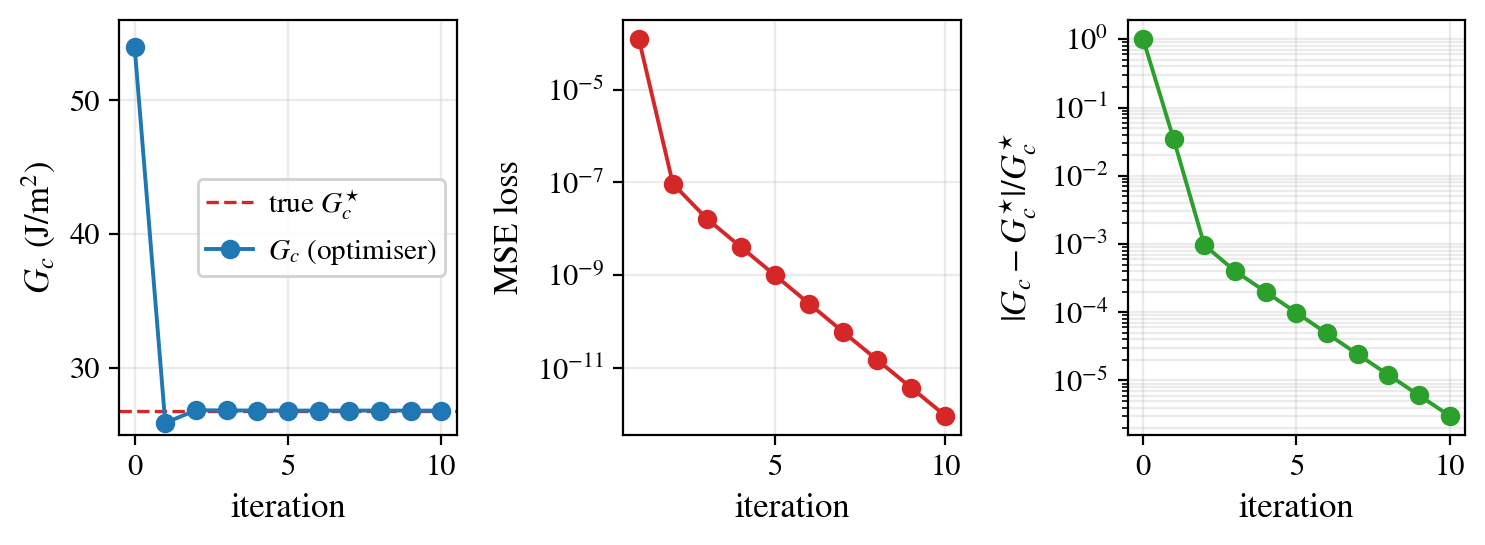}
  \caption{Autograd-driven L-BFGS material inversion on alumina
    (Al$_2$O$_3$) using the
    Kumar et al.~\cite{kumar2020revisiting} parameters,
    $\GC^{*} = 26.8$~J/m$^2$, $\GC^{(0)} = 54.0$~J/m$^2$ ($2\times$
    initial guess). The optimisation uses the same reverse-mode
    autograd gradient path as Figure~\ref{fig:inversion}. Only the
    material parameters change.}
  \label{fig:inversion_alumina}
\end{figure}

\subsection{Implicit differentiation through the CG damage solve}
\label{sec:implicit_diff}

The damage equation is solved by conjugate gradient iterations at every
time step. If reverse-mode autograd is applied directly, the backward
graph stores the full Krylov trajectory. Every matrix--vector product,
inner product, and scalar update in the CG solver becomes part of the
graph. For a solve requiring 20 CG iterations, the graph must retain the
intermediate Krylov vectors and scalar updates from all 20 iterations.
The memory cost therefore grows with the number of CG iterations and can
become prohibitive on fine meshes.

The implementation avoids this by differentiating through the converged
linear system instead of through the unrolled CG iterations. The forward
pass solves
\begin{equation}
  A(\theta)d = b(\theta)
  \label{eq:implicit_forward_linear_system}
\end{equation}
and returns the converged damage field $d$. The backward pass solves one
adjoint linear system with the same matrix-free operator. If $g$ is the
upstream gradient on $d$, the adjoint variable $\lambda$ is obtained
from
\begin{equation}
  A^{\top}\lambda = g .
  \label{eq:implicit_adjoint_linear_system}
\end{equation}
The parameter gradient is then computed from the dependence of
$A(\theta)$ and $b(\theta)$ on the material parameters.

This rule is implemented as a custom PyTorch autograd function for the
scalar toughness recovery. The forward pass runs the existing CG solver
with autograd disabled. The backward pass performs one additional CG
solve against the adjoint operator. For scalar parameters such as
$(\GC,\lz)$, the AT2 gradient on linear triangles reduces to element-wise
mass and stiffness contributions, for example
\begin{equation}
  -\left[\lz\,\lambda^{\top}\bs{K}\,d
  + (1/\lz)\,\lambda^{\top}\bs{M}\,d\right].
  \label{eq:implicit_gc_l0_element_term}
\end{equation}
These terms are computed by scatter operations without assembling a
sparse matrix.

The irreversibility constraint is handled by active-set masking. At
nodes clamped by $d \geq d_{\text{prev}}$, the upstream gradient passes
through the previous damage state. At interior nodes, the gradient flows
through the linear system. The backward rule is therefore a local
linearisation at the active set realised by the forward simulation. If
crack activation changes that active set, the sensitivity can change
abruptly.

The scalar recovery examples include independent finite-difference
checks at the initial iterate. The glass example agrees to $0.19\%$.
The alumina scale check agrees to $5.1\%$ on the realised active-set
path. These checks support the implemented derivative for the scalar
recoveries reported here. Larger inverse fracture problems will require
additional active-set robustness tests.

\section{Performance and comparison}
\label{sec:performance}

Smaller CPU tests and baseline runs were performed on an Apple M4 Pro
processor with 14 arm64 cores (10 performance and 4 efficiency cores) and
PyTorch~2.8.0 in 64-bit precision. The controlled CPU/GPU
comparison below used an Intel Windows Subsystem for Linux 2 (WSL2)
workstation with an NVIDIA RTX
A2000 (6~GB video random access memory, VRAM) and Compute Unified
Device Architecture (CUDA)~12.x. The large-scale benchmarks reported in
Section~\ref{sec:benchmarks} were run on NVIDIA A100 80~GB PCIe GPUs
with PyTorch~2.8.0 and CUDA~12.8; the A100 provides $9.7$~tera
floating-point operations per second (TFLOPS) of standard 64-bit
floating-point (FP64) throughput and substantially higher memory bandwidth than
the workstation A2000.

\subsection{Single-solver cost scaling}
\label{sec:a100_timing}

\begin{figure}[htb]
  \centering
  \includegraphics[width=\textwidth]{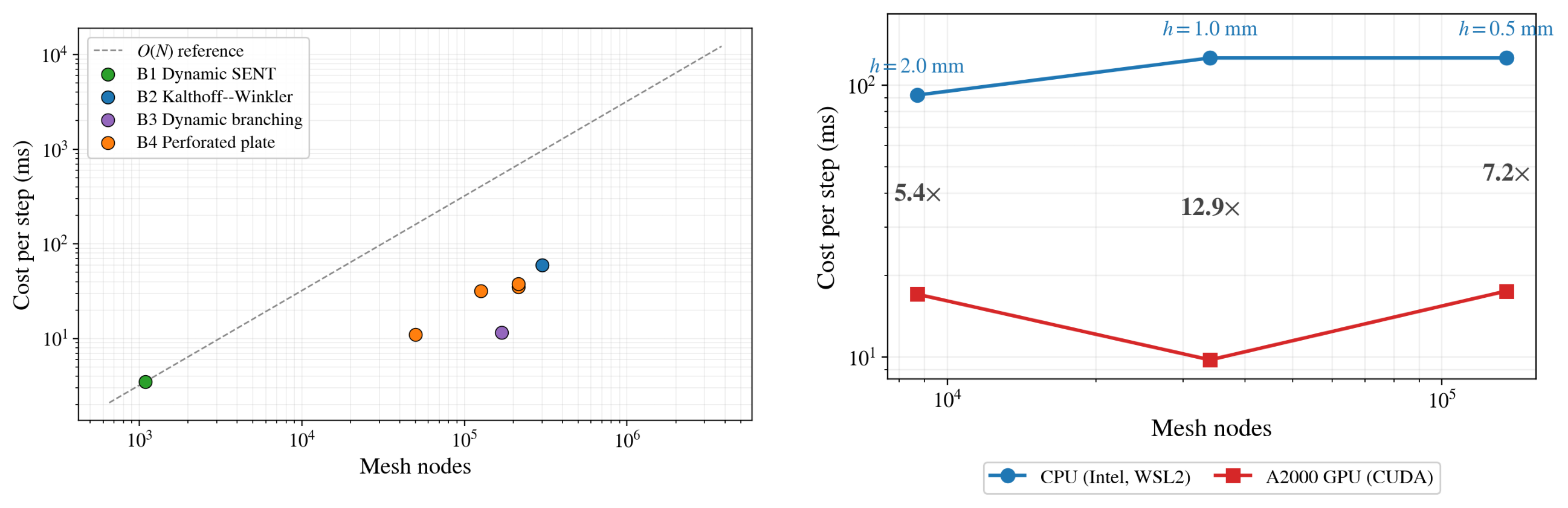}
  \caption{Performance scaling. (a) Measured wall-clock cost per explicit step
  on NVIDIA A100 80~GB for representative benchmark runs. Points are coloured
  by benchmark family; the dashed line anchors an $O(N)$ reference at the
  smallest run (B1 SENT, 1091 nodes, 3.5~ms/step). The measured points are
  consistent with near-linear scaling over this range, as expected when
  matrix-free element-wise kernels dominate the step cost. B3 is the dynamic crack-branching run
  ($169{,}077$ nodes and $336{,}266$ elements). (b) Controlled CPU vs
  A2000 GPU comparison on Kalthoff--Winkler, three mesh resolutions
  ($h = 2$, $1$, $0.5$~mm; $8{,}729$ to $134{,}961$ nodes), $5~\mu$s of
  simulated dynamics, Jacobi-preconditioned CG, double precision, identical
  problem setup. The workstation uses WSL2 Ubuntu, an Intel CPU, and an NVIDIA RTX A2000
  (6~GiB VRAM), CUDA~12.8, PyTorch~2.8. The GPU speedup is $5.4\times$ at the
  coarsest mesh, peaks at $12.9\times$ at the $34$k-node midpoint, and settles
  at $7.2\times$ at the finest mesh.}
  \label{fig:performance_scaling}
\end{figure}

Figure~\ref{fig:performance_scaling}(a) plots the measured wall-clock cost per
explicit time step on the A100 GPU for representative runs from each
benchmark, recorded by the solver's per-run metadata logging. The cost
is consistent with near-linear scaling over the measured range, as
expected for a matrix-free formulation in which the dominant element
kernels are $O(N_e)$.
The dynamic crack-branching run ($169{,}077$ nodes and
$336{,}266$ elements) completes the full $183{,}941$-step simulation in
approximately $36$ minutes of A100 wall-clock time, including benchmark
output and post-processing data written by the run; this corresponds to
a sustained $\sim\!11.6$~ms/step ($\sim\!1.45 \times 10^{7}$
node-updates per second, equivalently $\sim\!2.89 \times 10^{7}$
element-updates per second).

To isolate the hardware speedup from confounding effects introduced by comparing
an arm64 Apple CPU against a datacentre A100, timings were also measured for the
same Kalthoff--Winkler problem on CPU and on an NVIDIA RTX A2000 within
the same WSL2 workstation (same Python, same PyTorch build, same mesh
file, Jacobi-preconditioned CG, double precision).
Figure~\ref{fig:performance_scaling}(b) reports the result across three mesh
resolutions. At $34$k nodes the workstation GPU speedup reaches
$12.9\times$; the CPU rises from $91$ to $125$~ms/step between the
coarsest and middle mesh and then changes more slowly at the finest
mesh, behaviour consistent with cache and memory effects on the CPU
path. At the finest mesh the measured GPU speedup is $7.2\times$.
Peak GPU allocation for the $135$k-node run is $219$~MiB, an order of
magnitude below the A2000's $6$~GiB capacity. The timing is therefore
not limited by device capacity. Further profiling would be needed to
separate compute throughput from memory-bandwidth effects. As a
reproducibility check, the reported final damage-field extrema match
between CPU and GPU to machine precision.

A timed Kalthoff--Winkler run on $34$k nodes / $67$k elements,
integrated over the full $\sim\!20\,\mu$s simulated dynamics (562
explicit steps, CPU double precision, Jacobi-preconditioned CG),
decomposes approximately into mutually exclusive categories. The
damage CG solve accounts for $\sim\!59\%$ of the step cost, the
velocity-Verlet mechanics update (exclusive of the nested strain
computation) for $\sim\!19\%$, strain computation for $\sim\!16\%$,
the $\psi^{+}$ and history-variable update (exclusive of strain)
for $\sim\!6\%$, and Python/I-O bookkeeping for the remaining
$\sim\!0.2\%$ (rounded individually; total within the rounding
tolerance of $\pm 0.5\%$).

The damage fraction is higher on smaller meshes
(dominated by CG overhead) and decreases on larger meshes where the
$O(N_e)$ mechanics kernels take a larger share. The CG iteration
count itself is low throughout. It peaks at $\sim 7$ to $15$ when
damage first develops and drops to $1$ to $5$ during steady propagation,
because the previous damage field provides an increasingly accurate
initial guess. The subcycling savings scale with the damage-solve
fraction of the per-step cost, so the reported reduction is
mesh-dependent. With phase-field subcycling enabled at $N_{\text{sub}}=3$
(Section~\ref{sec:subcycling}) the measured wall-clock reduction is
$\sim 45\%$ on a $15{,}408$-node Kalthoff configuration where the
damage share is $\sim 71\%$, and is correspondingly smaller on the
$34{,}060$-node breakdown above where the damage share is $59\%$.

Memory consumption is modest for the core explicit state. The stored
state variables (displacement, velocity, acceleration, damage, history,
external forces, and per-element reference geometry) are of the same
order as a single sparse mechanics-stiffness matrix in
compressed-sparse-row (CSR) format with $\sim 14$ non-zeros per row
($=2\cdot7$ for linear triangles in two dimensions, as measured on the
benchmark meshes). The matrix-free mechanics path avoids the assembly
cost of that matrix. The optional algebraic-multigrid (AMG) damage
preconditioner dominates the additional storage when enabled, because
the hierarchy restriction and prolongation operators account for the
bulk of node-shaped storage at the benchmark resolutions. Switching to
Jacobi or unpreconditioned CG removes this cost at the price of a few
extra CG iterations at the onset of damage.

This is a familiar trade-off in the Krylov literature:
matrix-based methods expose incomplete LU (ILU), sparse algebraic
multigrid, and direct factorisation immediately, whereas matrix-free
methods usually need specialised
preconditioners such as geometric multigrid, $p$-multigrid, or
low-order surrogate matrices to recover competitive convergence
rates~\cite{austin2012sparseprecond,davydov2020matrixfreehyperelastic,
brown2022matrixfreepmg}.
\label{sec:cost_breakdown}

\subsection{Capability comparison with related frameworks}
\label{sec:comparison}

Each established framework occupies a distinct design point.
deal.II~\cite{bangerth2007deal} provides mature high-order
finite-element infrastructure, including matrix-free operator
evaluation for implicit solvers.
JAX-FEM~\cite{xue2023jax} delivers GPU-capable differentiable
assembly with adjoint-based vector--Jacobian products (VJPs). FEniCS-based
implementations~\cite{logg2012automated,alnaes2015fenics} combine a
mature variational interface with the dolfin-adjoint differentiation
toolchain. COMSOL Multiphysics~6.4 brings phase-field
fracture to commercial multiphysics environments and adds GPU
sparse-direct acceleration via cuDSS. The present solver targets a
complementary regime with explicit dynamics, no sparse
mechanics-stiffness assembly in the explicit update, and
autograd-compatible tensor operations on
commodity GPUs.

The distinguishing features of the present work are the combination of
three properties. First, it performs explicit dynamics on unstructured
meshes without assembling a sparse mechanics stiffness matrix in the
explicit update, keeping the mechanics operators matrix-free via
scatter-based accumulation. Second,
it provides GPU support without requiring specialised linear solver
backends. FEniCSx and JAX-FEM can run on GPUs, but their performance
paths typically still depend on sparse-matrix kernels or external
linear-algebra backends, whereas the present solver inherits PyTorch's
native CUDA backend with no compiled
extensions. Third, it exposes an autograd-compatible tensor path.
Adjoint-based frameworks (dolfin-adjoint, JAX-FEM~\cite{xue2023jax})
compose well with their assembled-matrix backbones via per-operation
VJPs; the matrix-free PyTorch design exposes the forward operators
directly to autograd, so user-added operations can inherit their
backward pass from the framework. The PyTorch design also lets the user
compose downstream neural-network losses readily.

\section{Conclusions}
\label{sec:conclusions}

A matrix-free PyTorch solver for phase-field fracture has been
presented, with differentiable tensor kernels and an
implicit-differentiation rule for the CG damage solve. The
matrix-free assembly removes sparse-matrix operations from the
explicit time-stepping loop. The same implementation runs unmodified on
macOS, Linux, and Windows, on both CPU and NVIDIA GPU backends, and
reaches mesh sizes of order $10^{6}$ nodes on a single A100. The
source repository provides the benchmark inputs, generated figures, and
auxiliary timing logs used for reproducibility checks. The forward
solver is compared against four dynamic
benchmarks (straight mode-I propagation, oblique shear-induced
kinking, dynamic crack branching, and crack-hole interaction)
and two quasi-static benchmarks (Miehe SENT and the
notched-holed plate). The benchmark comparisons, including the observed
agreement and remaining differences relative to the reference sources,
support the implementation across dynamic and quasi-static settings. The
quasi-static checks further show that the same formulation can be used
with the portable matrix-free CG solver. For the parameter-matched
implicit mechanics comparisons, the same formulation can also use
sparse-direct linear solvers. The
autograd-compatible explicit kernels and the constant-memory
implicit-differentiation rule for the CG damage solve provide the
gradient infrastructure for higher-dimensional inverse and design
studies, including neural-network training whose loss can be evaluated
through the solver.

\noindent The contributions can be summarised as follows.

\begin{itemize}
  \item A matrix-free, scatter-based assembly of the explicit dynamic
        phase-field operators. The explicit tensor kernels are
        compatible with PyTorch automatic differentiation, and the CG
        damage solve is wrapped in a constant-memory
        implicit-differentiation rule.
  \item Cross-code and cross-regime benchmark evidence on four dynamic
        fracture benchmarks from published phase-field studies and COMSOL application
        examples, two quasi-static
        benchmarks, and qualitative shared-mesh damage comparisons
        against FEniCS and Akantu.
  \item One demonstrated differentiability application is an L-BFGS
        scalar fracture-toughness inversion. Recovery of $\GC$ with
        relative error below $10^{-3}$ is achieved after only three
        accepted L-BFGS states for glass and two for alumina.
        Further optimiser iterations are retained as a stability check
        and confirm that the recovered value remains on a stable plateau. The
        autograd path through the CG damage solve provides a route to
        higher-dimensional inverse applications.
\end{itemize}

Several concrete extensions follow naturally from the present design.
Gradients through hundreds of time steps accumulate the full
activation tape in GPU memory, which bounds the product of mesh
size and step count addressable in a single run. Gradient
checkpointing~\cite{chen2016checkpointing} around the time-stepping
loop reduces this tape from $O(N_\mathrm{steps})$ to
$O(\sqrt{N_\mathrm{steps}})$ at the cost of one additional forward
pass, which is a step toward per-element $\GC(\bs{x})$ recovery on
larger meshes. Extensions to three-dimensional geometries and
adaptive mesh refinement remain natural next steps for the same
architecture.

\section*{Declaration of competing interests}

The authors declare that they have no known competing financial interests
or personal relationships that could have appeared to influence the work
reported in this paper.

\section*{Acknowledgements}

The authors gratefully acknowledge funding support from UK Research and
Innovation through Engineering and Physical Sciences Research Council
grant EP/Y004671/1, and computational resources provided by
City St~George's, University of London, including access to the
Department of Engineering high-performance computing facility and the
NVIDIA A100 GPUs used for the large-scale benchmark runs. The authors
also thank Shad Ali Durussel of EPFL's Computational Solid Mechanics
Laboratory and the Akantu team for helpful discussions on Akantu's
staggered-scheme treatment of phase-field fracture problems.

\section*{Data and code availability}

The solver, benchmark input scripts, mesh generators, and the
post-processing utilities required for the reported public figures
are openly available at
\url{https://github.com/CEMS-Lab/PhAST}. The explicit dynamic pathway is a
portable, matrix-free PyTorch implementation requiring only
Python~3.9+, PyTorch~2.0+, and Gmsh. For quasi-static verification the
same codebase can either keep this matrix-free CG route or opt into
available sparse-direct backends such as SciPy, the Portable,
Extensible Toolkit for Scientific Computation (PETSc), or the
MUltifrontal Massively Parallel sparse direct Solver (MUMPS) for
parameter-matched comparison studies. The repository also contains the
auxiliary timing logs and comparison notes used to audit performance
during development.

\bibliographystyle{elsarticle-num}
\bibliography{references}

\end{document}